\documentclass[12pt]{iopart}

\expandafter\let\csname equation*\endcsname\relax
\expandafter\let\csname endequation*\endcsname\relax

\usepackage[utf8]{inputenc}
\usepackage{amsmath}
\usepackage{amsfonts}
\usepackage{amssymb}
\usepackage{soul}
\usepackage{amsthm}
\usepackage[colorlinks=true]{hyperref}
\usepackage[normalem]{ulem}
\usepackage{bm}
\usepackage[T1]{fontenc}
\usepackage{graphicx}
\usepackage{subcaption}
\usepackage{caption}
\usepackage{subcaption}
\usepackage{relsize}
\usepackage{mathrsfs}
\DeclareUnicodeCharacter{2212}{-}
\usepackage[usenames,dvipsnames]{xcolor}

\makeatletter
\renewcommand\footnoterule{%
  \kern-3\p@
  \hrule\@width2.5cm
  \kern2.6\p@}
\makeatother

 \def\be{\begin{equation}}
\def\ee{\end{equation}}
 \def\ba{\begin{align}}
\def\ea{\end{align}}
\def\bea{\begin{eqnarray}}
\def\eea{\end{eqnarray}}
\def\d{\partial}

\def\a{\alpha}

\def\m{\mu}
\def\n{\nu}

\def\th{\theta}

\def\s{\sigma}

\def\P{{\cal P}}

\newcommand{\interior}{\mbox{\tiny int}}
\newcommand{\ext}{\mbox{\tiny ext}}
\newcommand{\glo}{\mbox{\tiny glo}}

\newcommand{\bv}{{\bm v}}

\newcommand{\bseq}{\begin{subequations}}
\newcommand{\eseq}{\end{subequations}}

\begin{document}

\vspace{-1cm}
\title[Updated Binary Pulsar Constraints on Einstein-\ae ther theory]{New Binary Pulsar Constraints on Einstein-{\ae}ther Theory after GW170817}

\author{Toral Gupta$^1$, Mario Herrero-Valea$^{2,3}$, Diego Blas$^4$, Enrico Barausse$^{2,3}$, Neil Cornish$^1$, Kent Yagi$^5$ and Nicol\'as Yunes$^6$}

\address{$^1$ eXtreme Gravity Institute, Department of Physics, Montana State University, Bozeman, Montana 59717, USA.}
\address{$^2$ SISSA, Via Bonomea 265, 34136 Trieste, Italy \& INFN, Sezione di Trieste.} 
\address{$^3$ IFPU - Institute for Fundamental Physics of the Universe, \\Via Beirut 2, 34014 Trieste, Italy.} 
\address{$^4$ Theoretical Particle Physics and Cosmology Group, Department of Physics, King's College London, Strand, London WC2R 2LS, UK.}
\address{$^5$ Department of Physics, University of Virginia, Charlottesville, Virginia 22904, USA.}
\address{$^6$ Illinois Center for Advanced Studies of the Universe, Department of Physics, University of Illinois at Urbana-Champaign, Urbana, Illinois, 61820 USA.}

\ead{toralgupta@montana.edu}

\date{today}

\begin{abstract}
The timing of millisecond pulsars has long been used as an exquisitely 
precise tool for testing the building blocks of general relativity, including
the strong equivalence principle and Lorentz symmetry.
Observations of binary systems involving at least one millisecond pulsar have been used to place bounds on the parameters of Einstein-\ae ther theory, a gravitational theory that violates Lorentz symmetry at low energies via a preferred and dynamical time threading of the spacetime manifold. 
However, these studies did not cover the region of parameter space that is still viable after the recent bounds on the speed of gravitational waves from  GW170817/GRB170817A. The restricted coverage was due to limitations in the methods used to compute the pulsar ``sensitivities'', which parameterize violations of the strong-equivalence principle in these systems. We extend here the calculation of pulsar sensitivities
to the parameter space of  Einstein-\ae ther theory that remains viable after GW170817/GRB170817A. We show that observations of the damping of the period of quasi-circular binary pulsars and of the triple system PSR J0337+1715 further constrain the viable parameter space by about an order of magnitude over previous constraints. 
\end{abstract} 
\vspace{-1cm}
\submitto{\CQG}

\maketitle

\section{Introduction}

Lorentz symmetry has been the foundation of the magnificent edifice of 
 theoretical physics for more than a century, playing a central role in special and general relativity (GR), as well as in the quantum theory of fields. Because of its special status, Lorentz invariance has been tested to exquisite precision in the  matter sector via particle physics experiments~\cite{kostelecky:2003fs,kostelecky:2008ts,Mattingly:2005re,Jacobson:2005bg}. More recently, this experimental program has been extended to the matter-gravity~\cite{kostelecky:2010ze},
 dark matter \cite{Blas:2012vn,Bettoni:2017lxf}, and pure-gravity sectors~\cite{Jacobson:2008aj,liberati:2013xla}, where bounds on Lorentz violations (LVs) have been historically looser (because of the intrinsic weakness of the gravitational interaction).
 
 Compelling theoretical reasons to  seriously consider the possibility of LVs in the purely gravitational sector were provided by the realization that
 they could generate a better behavior in the ultraviolet (UV) limit. In particular, P.~Ho\v rava~\cite{horava:2009uw} showed that by allowing for a non-isotropic scaling between space and time,
 one can construct a theory that is power-counting renormalizable in 
 the UV. Renormalizability beyond power counting (i.e.~pertubative renormalizability) 
 in special (``projectable'') versions of Ho\v rava gravity has also been proven~\cite{barvinsky:2015kil}. 
 
The low-energy limit of Ho\v rava gravity reduces to ``khronometric  theory''~\cite{Blas:2009qj,Jacobson:2010mx}, which consists of GR plus an additional hypersurface-orthogonal and timelike vector field, often referred to as the ``\ae ther''. Because this vector field is hypersurface orthogonal, it selects a preferred spacetime foliation, which makes LVs manifest. 
 A more general 
 boost-violating low-energy gravitational theory, however, can be obtained by relaxing the assumption that the \ae ther be hypersurface-orthogonal, in which case it selects a preferred time \textit{threading} of the spacetime rather than a preferred foliation. The resulting theory is known as Einstein-\ae ther theory~\cite{Jacobson:2000xp}.
 
Despite allowing for an improved UV behavior, LVs in gravity 
face long-standing experimental challenges, particularly
when it comes to their percolation into
the matter sector, where particle 
physics experiments are in excellent agreement with Lorentz symmetry.
While some degree of percolation is inevitable, because of the coupling
between matter and gravity, mechanisms suppressing it have been put forward, including suppression by a large energy scale~\cite{Pospelov:2010mp}, or
the effective emergence of Lorentz symmetry at low energies 
as a result of renormalization group flows~\cite{Chadha:1982qq,Bednik:2013nxa,barvinsky:2017kob} 
 or  accidental symmetries~\cite{GrootNibbelink:2004za}.

At the same time, purely gravitational bounds on LVs are becoming increasingly 
compelling. The parameters (``coupling constants'') of both Einstein-\ae ther  and khronometric theory have been
historically constrained by theoretical considerations (absence of ghosts and gradient instabilities~\cite{Blas:2011zd,Jacobson:2004ts,Garfinkle:2011iw}, well-posedness of the Cauchy problem~\cite{Sarbach:2019yso}), by the absence of vacuum Cherenkov cascades in cosmic-ray experiments~\cite{Elliott:2005va}), by solar-system tests~\cite{Will:2014kxa,foster:2005dk,Blas:2011zd,Bonetti:2015oda,muller:2005sr}, 
by observations of the primordial abundances of elements from Big-Bang nucleosynthesis~\cite{Carroll:2004ai}, by other cosmological tests \cite{Audren:2014hza}, and by 
precision timing of binary pulsars (where LVs generically predict
violations of the strong equivalence principle)~\cite{foster:2007gr,Foster:2006az,Yagi:2013ava,Yagi:2013qpa,Barausse:2019yuk}. More recently, the coincident detection~\cite{GBM:2017lvd,PhysRevLett.119.161101} of gravitational waves (GW170817) and gamma rays (GRB170817A) emitted by the coalescence of two neutron stars and the subsequent kilonova explosion has allowed extremely strong constraints on the propagation speed of gravitational waves, which must equal that of light to within\footnote{See also \cite{Cornish:2017jml} for looser bounds coming from mergers of black holes.} $10^{-15}$~\cite{Monitor:2017mdv}, which in turn places even more stringent bounds on the couplings of both theories~\cite{Gumrukcuoglu:2017ijh,ramos,Sarbach:2019yso,Oost:2018tcv}.

The bounds from the coincident GW170817/GRB170817A observations force us to rethink the parameter spaces of both Einstein-\ae ther  and khronometric theory, as the only currently allowed regions
appear to be ones that were previously thought to be of little interest, and which
were not explored extensively. In the case of khronometric theory, Refs.~\cite{ramos,Barausse:2019yuk} found that the  couplings that remain viable after GW170817 and GRB170817  
produce exactly no deviations away from the predictions of GR, not only in the solar system, but also in binary systems of compact objects, be they black holes (BHs) or neutron stars (NSs), to leading post-Newtonian (PN) order. Reference~\cite{Franchini:2021bpt} extended this
result to the quasinormal modes of spherically symmetric black holes and to fully non-linear (spherical) gravitational collapse, where again no deviations
from the GR predictions are found.
It would therefore seem that the most promising avenue to further test khronometric theory may be provided by cosmological observables (e.g. Big-Bang nucleosynthesis abundances or CMB physics), where the viable couplings do produce non-vanishing deviations away from the GR phenomenology. 

Like for khronometric theory, 
the parameter space where detailed predictions for isolated/binary pulsars
were obtained in Einstein-\ae ther theory~\cite{Yagi:2013ava,Yagi:2013qpa} does \textit{not} include the
region singled out by the combination of the GW170817/GRB170817A bound and existing solar-system constraints (see Ref.~\cite{Sarbach:2019yso} for a discussion). The goal of this paper is therefore to extend the previous analysis of binary/isolated-pulsar data by some of us~\cite{Yagi:2013ava,Yagi:2013qpa} to this region of parameter space. This will require a significant modification of the formalism that Refs.~\cite{Yagi:2013ava,Yagi:2013qpa} utilized to calculate 
pulsar ``sensitivities'', i.e.~the parameters that quantify  violations
of the strong-equivalence principle in these systems.
Moreover, we will extend our analysis to include additional data over that considered in Refs.~\cite{Yagi:2013ava,Yagi:2013qpa}, namely the triple system
 PSR J0337+1715~\cite{archibald:2018oxs}. 
 Overall, we find that observations of the damping of the period of quasi-circular binary pulsars, and that of the triple system PSR J0337+1715, reduce
the viable parameter space of Einstein-\ae ther theory by about an order of magnitude over previous constraints.
 
We will also amend an error (originally pointed out in Ref.~\cite{will:2018ont}) in the calculation of the strong-field preferred-frame parameters $\hat{\alpha}_1$ and $\hat{\alpha}_2$
for isolated pulsars, which were presented in Refs.~\cite{Yagi:2013ava,Yagi:2013qpa}. While we have checked that this error does not impact the bounds presented in Refs.~\cite{Yagi:2013ava,Yagi:2013qpa}, 
we present in \ref{app:EIH} a detailed derivation of $\hat{\alpha}_1$ and $\hat{\alpha}_2$ for possible future applications, also correcting a few typos present in the original calculation of Ref.~\cite{will:2018ont}.

This paper is organized as follows. In Sec.~\ref{sec:AE1} we give a succinct introduction to Einstein--\ae ther theory, including the modified field equations
and the current observational bounds on the coupling constants. In Sec.~\ref{sec:AE2} we
introduce the concept of stellar sensitivities as parameters regulating violations of the strong equivalence principle.
Solutions
describing slowly moving stars are derived in Sec.~\ref{sec:exp_V}, and they are used
in 
 Sec.~\ref{sec:sensit} to compute the
  sensitivities. Section \ref{sec:constr} uses the sensitivities to obtain the constraints on Einstein--\ae ther theory resulting from observations of binary and triple pulsar systems. We summarize our conclusions in Sec.~\ref{sec:concl}.
  \ref{app:EIH} contains a calculation of the strong-field preferred-frame parameters $\hat{\alpha}_1$ and $\hat{\alpha}_2$ in Einstein-\ae ther theory, fixing an oversight in~\cite{Yagi:2013ava}, which was pointed out by \cite{will:2018ont}, and correcting also a few typos present in \cite{will:2018ont} itself.
  We will adopt units where $c=1$ and a signature $+---$, in accordance with most of the literature on Einstein-\ae ther theory.

\section{Einstein \ae ther theory}\label{sec:AE1}

In order to break boost (and thus Lorentz) symmetry, 
Einstein-\ae ther theory introduces a dynamical threading
of the spacetime by a unit-norm, time-like vector field 
$\boldsymbol{U}$. This vector field, often referred to as the \ae ther, physically
represents a preferred ``time direction'' at each spacetime event. Requiring the action to also include the usual spin-2 graviton of GR, to be quadratic
in the \ae ther derivatives, and to feature no direct coupling between the matter and the \ae ther  (so as to enforce the weak equivalence principle, i.e. the universality of free fall, and the absence of matter LVs at tree level), one obtains the action~\cite{Jacobson:2000xp,jacobson_2014}
\begin{align}
S&=-\frac{1}{16\pi G}\int \Big[R + \frac13 c_\theta \theta^2 + c_\sigma  \sigma_{\mu\nu} \sigma^{\mu\nu}  + c_\omega \omega_{\mu\nu} \omega^{\mu\nu} + c_a A_\mu A^\mu\nonumber\\&+\lambda (U^\mu U_\mu-1)\Big]\sqrt{-g}\, d^{4}x+ S_{\rm mat}(\psi,g_{\mu\nu}),
\end{align}
where $R$ is the four-dimensional Ricci scalar, $g$ the determinant of the metric,
$G$ the bare gravitational constant (related to the value $G_N$ measured locally
by $G_N=G/(1-c_a/2)$~\cite{Carroll:2004ai,Jacobson:2008aj}), $\psi$ collectively denotes the matter degrees of freedom,
$\lambda$ is a Lagrange multiplier enforcing the \ae ther's unit norm,
$c_\theta$, $c_\sigma$, $c_\omega$ and $c_a$ are dimensionless constants\footnote{Note that much of the earlier literature on
Einstein-\ae ther theory uses a different set of coupling constants $c_i$ ($i=1,\ldots,4$), which are related to
our parameters by
$c_1 = (c_\omega + c_\sigma)/2$, 
$c_2 = (c_\theta - c_\sigma)/3$, 
$c_3 = (c_\sigma - c_\omega)/2$ and 
 $c_4 = c_a - (c_\sigma + c_\omega)/2$.
}, and we have decomposed the \ae ther congruence into the expansion $\theta$, the shear $\sigma_{\mu\nu}$, the vorticity $\omega_{\mu\nu}$
and the acceleration $A_\mu$ as follows:
\begin{gather}
A^\mu= U^\nu\nabla_\nu U^\mu\, ,\\ 
\theta=\nabla_\mu U^\mu\,,\\
\sigma_{\mu\nu}=\nabla_{(\nu}U_{\mu)}+A_{(\mu}U_{\nu)}-\frac13\theta h_{\mu\nu}\, ,\\
\omega_{\mu\nu}=\nabla_{[\nu}U_{\mu]}+A_{[\mu}U_{\nu]}\, ,
 \end{gather}
with $h_{\mu\nu}=g_{\mu\nu}-U_\mu U_\nu$ the projector onto the hyperspace orthogonal to $\boldsymbol{U}$. 

By varying the action with respect to the metric, the \ae ther and the Lagrange multiplier, and 
by eliminating the latter from the equations, one obtains the
generalized Einstein equations 
\begin{equation}\label{aether_ein_eq}
    E_{\alpha \beta} \equiv G_{\alpha \beta} - T^{\text{\AE}}_{\alpha \beta} - 8\pi G T_{\alpha \beta}^{\rm mat} = 0
\end{equation}
and the  \ae ther  equations 
\begin{equation}\label{aethereqn}
    \text{\AE}_{\mu} = \left[\nabla_{\alpha}J^{\alpha \nu} - \left(c_{a}-\frac{c_{\sigma}+c_{\omega}}{2}\right)A_{\alpha}\nabla^{\nu}U^{\alpha}\right] h_{\mu\nu} = 0 ,
\end{equation}
where $ G_{\alpha \beta }$ is the Einstein tensor, the \ae ther stress-energy tensor is 
\begin{align}\label{T_ae}
    \nonumber T_{\alpha \beta}^{\text{\AE}} &=  \nabla_{\mu}\left(J_{(\alpha}\phantom{}^{\mu}U_{\beta)} - J^{\mu}\phantom{}_{(\alpha}U_{\beta)} - J_{(\alpha \beta)}U^{\mu}\right) + \frac{c_{\omega}+c_{\sigma}}{2}\left[(\nabla_{\mu}U_{\alpha})(\nabla^{\mu}U_{\beta}) - (\nabla_{\alpha}U_{\mu})(\nabla_{\beta}U^{\mu})\right] \\
    & + U_{\nu}(\nabla_{\mu}J^{\mu \nu})U_{\alpha}U_{\beta} - \left(c_{a}-\frac{c_\sigma+c_\omega}{2}\right)\left[A^{2} U_{\alpha}U_{\beta} - A_{\alpha}{A_{\beta}}\right] + \frac{1}{2}M^{\s \rho} \phantom{}_{\mu \nu}\nabla_{\s}U^{\mu}\nabla_{\rho}U^{\nu}g_{\alpha \beta},
\end{align}
with 
\begin{align*}
    &J^{\alpha}\phantom{}_{\mu} \equiv M^{\alpha \beta}\phantom{}_{\mu \nu} \nabla_{\beta}U^{\nu},\\
    &M^{\alpha\beta}\phantom{}_{\m\n}= \left(\frac{c_\sigma+c_\omega}{2}\right) h^{\alpha\beta}g_{\m\n}+ \left(\frac{c_\theta -c_\sigma}{3}\right) \delta^\alpha_\m \delta^\beta_\n +\left(\frac{c_\sigma - c_\omega}{2}\right) \delta^\alpha_\n \delta^\beta_\m + c_{a} U^\alpha U^\beta g_{\m\n},
\end{align*}
and the matter stress-energy tensor is defined as usual by
\be{}
    T^{\alpha \beta}_{\rm mat} \equiv - \frac{2}{\sqrt{-g}}\frac{\delta S_{\rm mat}}{\delta g_{\alpha \beta}} .
\ee{}

As already mentioned, a  number of experimental and theoretical results constrain
Einstein-\ae ther theory and the couplings $c_i$. In more detail,
perturbing  the field equations about Minkowski space 
yields propagation equations for spin-0 (i.e. scalar), spin-1 (i.e. vector) and spin-2 (i.e. tensor gravitons) modes. Their propagation speeds are 
respectively given by~\cite{Jacobson:2004ts}
\begin{gather}
c_T^2=\frac{1}{1-c_{\sigma}}\, ,
\label{s2}\\
c_V^2=\frac{c_{\sigma}+ c_{\omega} -
c_{\sigma} c_{\omega}}{2c_{a}(1-c_{\sigma})} \, ,
\label{s1}\\
c_S^2=\frac{(c_{\theta}+2c_{\sigma})(1-c_{a}/2)}{3c_{a}(1-c_{\sigma})(1+c_{\theta}/2)}\, .
\label{s0}
\end{gather}
In order to ensure stability at the classical level (i.e. no gradient instabilities) and at the quantum level (i.e. no ghosts) one needs
to have 
 $c_T^2>0$, $c_V^2>0$ and $c_S^2>0$~\cite{Jacobson:2004ts,Garfinkle:2011iw}. If we also require the modes to carry positive energy, we get $c_a>0$ and $c_\omega>0$~\cite{eling:2005zq}.
Furthermore, significantly subluminal graviton propagation 
would cause ultrarelativistic matter
to lose  energy to gravitons via a Cherenkov-like process~\cite{Elliott:2005va}.
Since this effect is not observed e.g. in ultrahigh energy cosmic rays, one must have 
 $c_I^2\gtrsim1-{\cal O}(10^{-15})$ (with $I=T,V,S$).
 More recently, the coincident detection of a neutron-star merger in GW170817 (gravitational waves)
 and GRB170817A (gamma rays) had led to the bound $-3\times10^{-15} <c_T-1< 7\times10^{-16}$~\cite{Monitor:2017mdv}.
 
Expanding the field equations through 1PN order leads to the conclusion
that the 1PN dynamics is well described (like in GR) 
by the parametrized PN (PPN) expansion~\cite{Will:2014kxa,muller:2005sr}. However, unlike in GR,
the preferred frame parameters $\alpha_1$ and $\alpha_2$ appearing in the PPN expansion do not vanish, but are given by~\cite{foster:2005dk}
\begin{gather}
    \alpha_1= 4\frac{c_{\omega}(c_{a}-2c_{\sigma})+c_{a}c_{\sigma}}{c_{\omega}(c_{\sigma}-1)-c_{\sigma}}\, ,
\label{eq:alpha1}\\
    \alpha_2=\frac{\alpha_1}{2}
    +\frac{3(c_{a}-2c_{\sigma})(c_{\theta}+c_{a})}{(2-c_{a})(c_{\theta} + 2c_{\sigma})} \, .
\label{eq:alpha2}
\end{gather}


Solar system experiments require $|\alpha_1|\lesssim 10^{-4}$ and $|\alpha_2|\lesssim10^{-7}$~\cite{Will:2014kxa,muller:2005sr}. By \textit{saturating} these bounds (i.e.~requiring in particular that $|\alpha_1|\lesssim 10^{-4}$ but \textit{not} $|\alpha_1|\ll 10^{-4}$) and combining them with the 
 constraints on the propagation speeds, one finds 
 $c_\sigma\approx {\cal O}(10^{-15})$, $c_a\approx {\cal O}(10^{-4})$, and $c_\theta\approx 3 c_a [1+{\cal O}(10^{-3})]$.
 The resulting experimentally viable parameter space, therefore,
 is effectively (i.e. to within a fractional width of $10^{-4}$ or better in the  parameters) one-dimensional: 
 $c_\sigma, \,c_a,\, c_\theta \approx 0$,
 but $c_\omega$
is essentially unconstrained~\cite{Sarbach:2019yso}.

Another viable region of the parameter space can be obtained by 
\textit{not} saturating the PPN constraints~\cite{Sarbach:2019yso}. In more detail,
one may require  $|\alpha_1|$ be much smaller than its upper limit, so as to automatically satisfy the bound on $\alpha_2$ 
(since $\alpha_2\propto{\alpha_1}$ if $c_\sigma\approx 0$, as imposed by GW170817 and GRB170817A). This leads to
$|c_a|\lesssim 10^{-7}$ and thus to
an effectively  two-dimensional experimentally viable parameter  space $(c_\theta,c_\omega)$, with the only additional 
requirement that $|c_\theta|\lesssim 0.3$ to ensure that the production of light elements
during Big Bang Nucleosynthesis gives predictions in agreement with observations~\cite{Carroll:2004ai}.

Both of the viable regions of parameter space identified above were \textit{not} considered in  Refs.~\cite{Yagi:2013ava,Yagi:2013qpa}, where neutron-star sensitivities in Einstein-\ae ther theory were first computed. This is because back when Refs.~\cite{Yagi:2013ava,Yagi:2013qpa} were written, the strongest constraints available were the solar system ones (since the GW170817/GRB170817A constraint was not yet available).  Refs.~\cite{Yagi:2013ava,Yagi:2013qpa} solved for $c_a$ and $c_\theta$ in terms of $c_\sigma$,  $c_\omega$, $\alpha_1$ and $\alpha_2$, and fixed the latter two
to their largest allowed values (respectively
 $\alpha_1=10^{-4}$ and $\alpha_2= 10^{-7}$).
Therefore, \textit{(i)} this does not include the 
first region listed above [$(c_\theta,c_\sigma,c_a) \lesssim 10^{-4}$ with $c_\omega$ kept free], which is obtained by choosing $\alpha_1\propto \alpha_2$; and \textit{(ii)} 
Refs.~\cite{Yagi:2013ava,Yagi:2013qpa} did not sample accurately
enough the sub-region $c_\sigma\approx 0$, $\alpha_1\lesssim 10^{-4}$, $\alpha_2\lesssim 10^{-7}$, since there was no reason to do so at that time and because the techniques
employed broke down in that sub-region (as we will show in the following).

\section{Strong-equivalence principle violations and sensitivities}\label{sec:AE2}

Most theories extending/modifying GR  involve additional degrees of freedom besides the massless tensor graviton of GR. These additional gravitational polarizations cannot directly couple with matter significantly,  to avoid introducing unwanted fifth forces in particle physics experiments, and to prevent violations of the weak equivalence principle (and particularly violations of the universality of free fall for weakly gravitating objects). Nevertheless, effective couplings between the extra gravitons and matter may be mediated by the metric perturbations (i.e. by the tensor gravitons present also in GR), which are typically coupled non-minimally to the extra gravitational degrees of freedom. These effective couplings become important when the metric perturbations are ``large'', which is the case for strongly gravitating systems such as those involving NSs and/or BHs.

A useful way to parametrize this effective coupling is provided by the sensitivity parameters. Because of the aforementioned effective couplings, the mass of strongly gravitating objects  will be  comprised not only of the contributions from matter and the metric (like in GR), but  it will also generally depend on the additional gravitational fields. We can thus describe isolated objects, and members of a widely separated binary, by a point particle model (like in GR), but with a non-constant mass depending on the extra fields. Because the mass is a scalar quantity, it must depend on a scalar constructed from the \ae ther field $\boldsymbol{U}$, the simplest of which is the Lorentz factor $\gamma\equiv\boldsymbol{u}\cdot\boldsymbol{U}$, where $\boldsymbol{u}$ is the particle's (i.e. the body's) four-velocity. 

In many practical situations (including the long inspiral of a binary system of compact objects) one may assume that the relative speed between the \ae ther and the object is small compared to the speed of light, and thus Taylor-expand the mass $\mu(\gamma)$ around $\gamma=1$: 
\begin{equation}
    \mu(\gamma)=\tilde{m} \left[1+\sigma (1-\gamma)+\frac12 \sigma' (1-\gamma)^2+ \ldots\right]\,
\end{equation}
where $\tilde{m}$, $\sigma$ and $\sigma'$ are constant parameters. In particular, the latter two are often referred to as the ``sensitivities'' and their derivatives:
\begin{align}
\sigma&\equiv -\frac{{\rm d}\ln \mu(\gamma)}{{\rm d} \ln\gamma}\Big|_{\gamma=1}\,,
\\
\sigma'&\equiv \sigma+\sigma^2+\frac{{\rm d}^2\ln \mu(\gamma)}{{\rm d} (\ln\gamma)^2}\Big|_{\gamma=1}\,.
\end{align}

In order to understand the effect of the sensitivities and their derivatives on the dynamics of binary systems, one can derive the equations of motion simply by varying the point particle action
\begin{equation}
    S_{\rm pp}= - \sum_A \int\mu_A(\gamma_A) d\tau_A\,, 
\end{equation}
where $A$ is an index identifying the objects, and $\tau$ is the proper time.
This yields the equation of motion
\begin{equation}
  [\mu_A(\gamma_A)-\mu'_A(\gamma_A)\gamma_A]  a^A_\beta=-\mu_A'(\gamma_A) (-u_A^\mu \nabla_\beta U^A_\mu+ u_A^\mu\nabla_\mu U^A_\beta)-\mu_A''(\gamma_A) \dot{\gamma}_A (U^A_\beta-\gamma_A u^A_\beta)\,,
\end{equation}
where again the index $A$ identifies the particle
under consideration (when used in $\mu$, $\gamma$ and $\boldsymbol{u}$) or at which position the \ae ther field $\boldsymbol{U}$ and its acceleration $\boldsymbol{A}$
are to be computed, the prime denotes a derivative with respect to the function's argument, and the overdot represents a derivative along $\boldsymbol{u}$ (i.e.
with respect to the proper time).

Reinstating the dependence on the speed of light $c$
and expanding in PN orders (i.e. for $c\to\infty$), one obtains the 1PN equations of motion for a 
binary as 
\begin{eqnarray}
&\frac{{\rm d}\boldsymbol{v}_A}{dt} =  -\frac{m_B \bm{n}}{r^2} \left \{{\cal G}_{AB} 
- \left (3{\cal G}_{AB}{\cal B}_{AB} + {\cal D}_{ABB} \right ) \frac{m_B}{r}
\right .
\nonumber \\
& 
\left .
- \frac{1}{2} \left [ 2{\cal G}_{AB}^2 + 6{\cal G}_{AB}{\cal B}_{(AB)} + 2{\cal D}_{BAA} + {\cal G}_{AB}({\cal C}_{AB} + {\cal E}_{AB}) \right ] \frac{m_A}{r} 
\right .
\nonumber \\
& 
\left .
+ \frac{1}{2}  \left [3{\cal B}_{AB} - {\cal G}_{AB} (1+ {\cal A}_A) \right ] v_A^2
+ \frac{1}{2}(3{\cal B}_{21} + {\cal G}_{AB}+ {\cal E}_{AB}) v_B^2 
\right .
\nonumber \\
& 
\left .
- \frac{1}{2} \left ( 6{\cal B}_{(AB)} + 2 {\cal G}_{AB} + {\cal C}_{AB} + {\cal E}_{AB} \right ) \bm{v}_A \cdot \bm{v}_B  
- \frac{3}{2} \left ({\cal G}_{AB}+ {\cal E}_{AB} \right ) (\bm{n} \cdot \bv_B)^2 \Big \}
\right .\nonumber \\
& 
+  \frac{m_B \bm{v}_A}{r^2} \bm{n} \cdot \left \{ \left [3{\cal B}_{AB} + {\cal G}_{AB} (1+ {\cal A}_A) \right ] \bv_A - 3{\cal B}_{AB} \bv_B  \right \} 
\nonumber \\
&
- \frac{1}{2} \frac{m_B \bm{v}_B}{r^2} \bm{n} \cdot \left [
\left ( 6{\cal B}_{(AB)} + 2 {\cal G}_{AB} + {\cal C}_{AB} + {\cal E}_{AB} \right ) \bv_A
-\left ( 6{\cal B}_{(AB)}  + {\cal C}_{AB} - {\cal E}_{AB} \right ) \bv_B
 \right ] 
 \nonumber \\
& 
- \frac{1}{2} \frac{m_B \bm{w}}{r^2} \bm{n} \cdot \left [
\left (  {\cal C}_{AB} -6{\cal B}_{[AB]} + {\cal E}_{AB} -2{\cal G}_{AB}  {\cal A}_A \right ) \bv_A
- \left ({\cal C}_{AB} - 6{\cal B}_{[AB]} - {\cal E}_{AB}  \right ) \bv_B
 \right]
\,,\nonumber\\
\label{acc}
\end{eqnarray}
where indices $A\neq B$, $r\equiv |\boldsymbol{x}_A-\boldsymbol{x}_B|$, $\boldsymbol{n}\equiv (\boldsymbol{x}_A-\boldsymbol{x}_B)/r$, $\bv_A\equiv {\rm d} \boldsymbol{x}/{\rm d} t$, and
\begin{eqnarray}
{\cal G}_{AB} &= \frac{G_N}{(1+\sigma_A)(1+\sigma_B)} \,, \nonumber \\
 {\cal A}_A &=   - \frac{\sigma'_A}{1+\sigma_A} \,,\nonumber \\
{\cal B}_{AB} &= {\cal G}_{AB} (1+\sigma_A) \,,
\nonumber \\
{\cal D}_{ABB} &={{\cal G}_{AB}}^2 (1+\sigma_A) \,,
\nonumber \\
{\cal C}_{AB} &=  {\cal G}_{AB} \left [  \alpha_1 - \alpha_2 - 3 \left (\sigma_A + \sigma_B \right )  - {\cal Q}_{AB} -{\cal R}_{AB}  \right ] \,,
\nonumber \\
{\cal E}_{AB} &= {\cal G}_{AB} \left [  \alpha_2 + {\cal Q}_{AB} - {\cal R}_{AB}  \right ] \,,
\nonumber \\
{\cal Q}_{AB} &= -\frac{1}{2} \left (\frac{2-c_{a}}{2c_{\sigma} - c_{a}} \right ) (\alpha_1 - 2\alpha_2) (\sigma_A +\sigma_B) +  3\left( \frac{2 - c_{a}}{2c_\sigma+ c_\theta}\right) \sigma_A \sigma_B  \,,
\nonumber \\
{\cal R}_{AB} &= \frac{1}{2} \left( \frac{8+\alpha_1}{c_\omega + c_\sigma}\right) \left [ -c_{\omega} (\sigma_A + \sigma_B) + (1-c_{\omega}) \sigma_A\sigma_B \right ]\,.
\label{eq10:EIHcoeffs2}
\end{eqnarray}

Note  that we have defined the ``active'' masses $m_A\equiv \tilde{m}_A (1+\sigma_A)$, in terms of which the Newtonian acceleration matches the GR result, albeit with a rescaled gravitational constant ${\cal G}_{AB}$. To derive Eq.~\eqref{acc} we have also used the 
PN-expanded solutions for the metric and \ae ther found in~\cite{foster:2007gr,Yagi:2013ava} (dropping divergent terms due to the point particle approximation, as usual in PN calculations):
\begin{align}
\label{g00PN}
g_{00} &= 1 -\frac{2 G_{N} \tilde{m}_1}{r_1 c^2} + \frac{1}{c^{4}} \left[ \frac{2 G_{N}^{2} \tilde{m}_1^{2}}{r_1^{2}} + \frac{2 G_{N}^{2} \tilde{m}_1 \tilde{m}_2}{r_1 r_2} + \frac{2 G_{N}^{2} \tilde{m}_1 \tilde{m}_2}{r_1 r_{12}} - \frac{3 G_{N} \tilde{m}_1}{r_1} v_1^{2} \left(1 + \sigma_1\right) \right]\nonumber\\&+ 1 \leftrightarrow 2 + {\cal{O}}(1/c^{6})\,,
\\
\label{g0iPN}
g_{0i} &= -\frac{1}{c^{3}} \left[B_1^{-} \frac{G_{N} \tilde{m}_1}{r_1} v_1^{i} + B_1^{+} \frac{G_{N} \tilde{m}_1}{r_1} v_1^{j} n_1^{j}  n_1^{i} \right]+ 1 \leftrightarrow 2 + {\cal{O}}(1/c^{4})\,,
\\
\label{gijPN}
g_{ij} &= -\left(1 + \frac{1}{c^{2}} \frac{2 G_{N} \tilde{m}_1}{r_1}\right) \delta_{ij} + 1 \leftrightarrow 2 + {\cal{O}}(1/c^{4})\,,
    \\
\label{ut_1E}
U^{0} &= 1 + \frac{1}{c^2}\frac{G_{N} \tilde{m}_1}{r_1} + 1 \leftrightarrow 2 + {\cal{O}}(1/c^{4})\,,
 \\\label{ui_1E}
U^{i} &= \frac{1}{c^3} \frac{G_{N} \tilde{m}_1}{r_1} \left(C_1^{-}  v_1^{i} + C_1^{+} v_1^j  n_1^{j}  n_1^{i} \right)+ 1 \leftrightarrow 2 + {\cal{O}}(1/c^{5})\,,
\\
B^{\pm}_{A} &\equiv  \pm \frac{3}{2} -2 \pm \frac{1}{4} (\alpha_1 -
2\alpha_2) \left( 1 + \frac{2-c_{a}}{2 c_{\sigma}-c_{a}}\sigma_{A} \right)
 - \frac{2c_{\omega}}{c_\omega + c_\sigma} \sigma_{A} - \frac{1}{4} \alpha_1 \left( 1+ \frac{c_{\omega}}{c_\omega + c_\sigma} \sigma_{A} \right)\,,
\\
C^{\pm}_{A} & \equiv \frac{1}{4} \left(\frac{8+\alpha_1}{c_\omega + c_\sigma}\right) [ c_{\omega} - (1-c_{\omega})\sigma_{A} ]
 \pm \frac{2-c_{a}}{2} \left( \frac{2\alpha_2-\alpha_1}{2
  (2c_\sigma-c_a)} + \frac{3 \sigma_{A}}{2c_\sigma + c_\theta}  \right)\,,
\end{align}
where 
 $r_{A}\equiv|\boldsymbol{x}-\boldsymbol{x}_A|$ and $\boldsymbol{n}_A\equiv{(\boldsymbol{x}-\boldsymbol{x}_A)}/r_A$.

Note that the \ae ther solution \eqref{ut_1E}--\eqref{ui_1E}
has space components $U^i$ vanishing at large distances from the binary, i.e. the equations of motion are valid in a preferred reference frame in which the \ae ther is asymptotically at rest. The dependence of the dynamics on the velocity $\boldsymbol{w}$ of the binary's center of mass with respect to the preferred frame can be reinstated by performing a boost. If $w\ll c$, one then obtains~\cite{will:2018ont}
\begin{eqnarray}\label{eq:acceleration}
&\frac{{\rm d}\boldsymbol{v}_A}{dt} =  -\frac{m_B \bm{n}}{r^2} \left \{{\cal G}_{AB} 
- \left (3{\cal G}_{AB}{\cal B}_{AB} + {\cal D}_{ABB} \right ) \frac{m_B}{r}
\right .
\nonumber \\
& 
\left .
\quad \quad
- \frac{1}{2} \left [ 2{\cal G}_{AB}^2 + 6{\cal G}_{AB}{\cal B}_{(AB)} + 2{\cal D}_{BAA} + {\cal G}_{AB}({\cal C}_{AB} + {\cal E}_{AB}) \right ] \frac{m_A}{r} 
\right .
\nonumber \\
& 
\left .
\quad \quad
+ \frac{1}{2}  \left [3{\cal B}_{AB} - {\cal G}_{AB} (1+ {\cal A}_A) \right ] v_A^2
+ \frac{1}{2}(3{\cal B}_{21} + {\cal G}_{AB}+ {\cal E}_{AB}) v_B^2 
\right .
\nonumber \\
& 
\left .
\quad \quad
- \frac{1}{2} \left ( 6{\cal B}_{(AB)} + 2 {\cal G}_{AB} + {\cal C}_{AB} + {\cal E}_{AB} \right ) \bm{v}_A \cdot \bm{v}_B  
- \frac{3}{2} \left ({\cal G}_{AB}+ {\cal E}_{AB} \right ) (\bm{n} \cdot \bv_B)^2 
\right .
\nonumber \\
& 
\left .
\quad \quad
+ \frac{1}{2} \left ( {\cal C}_{AB} + {\cal G}_{AB}  {\cal A}_A \right ) w^2
+ \frac{1}{2} \left ( {\cal C}_{AB} - 6{\cal B}_{[AB]}+ {\cal E}_{AB} + 2 {\cal G}_{AB}  {\cal A}_A \right ) \bv_A \cdot \bm{w}
\right .
\nonumber \\
& 
\left .
\quad \quad
+ \frac{1}{2} \left ( {\cal C}_{AB} + 6{\cal B}_{[AB]}- {\cal E}_{AB}  \right ) \bv_B \cdot \bm{w}
\right .
\nonumber \\
& 
\left .
\quad \quad
+ \frac{3}{2} {\cal E}_{AB} \left [ (\bm{w} \cdot \bm{n} )^2 + 2(\bm{w} \cdot \bm{n} )(\bm{v}_B \cdot \bm{n} ) \right ]
\right \}
\nonumber \\
& \quad
+  \frac{m_B \bm{v}_A}{r^2} \bm{n} \cdot \left \{ \left [3{\cal B}_{AB} + {\cal G}_{AB} (1+ {\cal A}_A) \right ] \bv_A - 3{\cal B}_{AB} \bv_B  + {\cal G}_{AB}  {\cal A}_A \bm{w}  \right \} 
\nonumber \\
& \quad
- \frac{1}{2} \frac{m_B \bm{v}_B}{r^2} \bm{n} \cdot \left \{
\left ( 6{\cal B}_{(AB)} + 2 {\cal G}_{AB} + {\cal C}_{AB} + {\cal E}_{AB} \right ) \bv_A
\right .
\nonumber \\
& 
\left .
\quad \quad
-\left ( 6{\cal B}_{(AB)}  + {\cal C}_{AB} - {\cal E}_{AB} \right ) \bv_B
+2 {\cal E}_{AB} \bm{w} 
 \right \} 
 \nonumber \\
& \quad
- \frac{1}{2} \frac{m_B \bm{w}}{r^2} \bm{n} \cdot \left \{
\left (  {\cal C}_{AB} -6{\cal B}_{[AB]} + {\cal E}_{AB} -2{\cal G}_{AB}  {\cal A}_A \right ) \bv_A
\right .
\nonumber \\
& 
\left .
\quad \quad
- \left ({\cal C}_{AB} - 6{\cal B}_{[AB]} - {\cal E}_{AB}  \right ) \bv_B
-2 \left ( {\cal G}_{AB}  {\cal A}_A - {\cal E}_{AB}  \right ) \bm{w} 
 \right \} 
\,,
\nonumber \\
\label{eq10:2bodyeom}
\end{eqnarray}
with which  Eq.~\eqref{acc} of course agrees for $\boldsymbol{w}=0$.

The sensitivities and their derivatives enter into the conservative dynamics of the binary system at Newtonian and 1PN order, as can be seen explicitly in Eqs.~\eqref{acc} and \eqref{eq:acceleration}. 
In \ref{app:EIH}  we will use these equations as starting point for studying in more detail the 1PN dynamics of binaries in Einstein-\ae ther theory. In doing so, we will also amend the calculation of the strong-field preferred-frame parameters $\hat{\alpha}_1$ and $\hat{\alpha}_2$ performed in~\cite{Yagi:2013ava}, fixing an oversight  pointed out by \cite{will:2018ont} and correcting also a few typos present in \cite{will:2018ont} itself\footnote{The corrections to $\hat \alpha_1$
and $\hat \alpha_2$ do not significantly affect the final results of~\cite{Yagi:2013ava}, 
since the strong-field preferred-frame parameters did not play a crucial role in constraining the parameter space of Lorentz-violating gravity in that paper. The measurements of $\hat \alpha_1$
and $\hat \alpha_2$ were used mainly to constrain $c_\omega$, but similar bounds on that coupling constant can be obtained from the  measurement of the damping of the period of PSR J0348+0432 (see Fig.~7 of~\cite{Yagi:2013ava}).}.

The sensitivities and their derivatives, however, also enter the dissipative dynamics. In more detail
the total energy emitted in GWs (including not only tensor but also scalar and vector \ae ther modes) by 
a binary in quasi-circular orbits was derived in~\cite{foster:2007gr,Yagi:2013ava}
via a standard multipole expansion and reads
\begin{align}
\label{Pdot-AE}
\frac{\dot{E}_{b}}{E_{b}} &=  \displaystyle 2    \left(\frac{ {\cal G}_{12} G m_1 m_2}{r^3}\right) 
  \Bigg\{ \frac{32}{5}({\Psi}_1+{\cal S}{\Psi}_2+ {\cal S}^2{\Psi}_3) v_{21}^2 \nonumber \\ 
  &\displaystyle+\left(s_1-s_2\right)^2\Bigg[{\zeta}_2+2{\zeta}_3 [w^2-(\boldsymbol{w}\cdot\boldsymbol{n})^2]+\frac{18}{5}{\Psi}_3\, w^2+\left(\frac{6}{5}{\Psi}_3+36 {\zeta}_1\right) (\boldsymbol{w}\cdot\boldsymbol{n})^2\Bigg] 
 \Bigg\}\,,
\end{align}
where $s_A\equiv{\sigma_A}/({1+\sigma_A})$ is the rescaled sensitivity for the $A$-th body; $\boldsymbol{v}_{21}=\boldsymbol{v}_{2}-\boldsymbol{v}_{1}$ is the relative velocity of the two bodies; 
the total (potential and kinetic) energy of the binary  is
\be
E_{b} = - \frac{{\cal{G}}_{12} m_1 m_2}{2 r}\,;
\ee
$w$ is the velocity of the binary's center of mass 
with respect to 
the preferred frame; and we
have introduced the definitions 
\begin{align}
\label{A-funcs-1}
{\Psi}_1 &\equiv \frac{1}{c_T}+\frac{2 c_{a} c_{\sigma}^2}{(c_\sigma + c_\omega-c_{\sigma} c_{\omega})^2 c_V}+\frac{3 c_{a}(Z-1)^2}{2 c_S(2-c_{a})}\,,  
\\ 
{\Psi}_2 &\equiv \frac{2(Z-1)}{(c_{a}-2)c_S^3} - \frac{2 c_{\sigma}}{(c_\sigma + c_\omega-c_{\sigma}c_{\omega})c_V^3}\,,  
\\ 
\label{A-funcs-3}
{\Psi}_3 &\equiv\frac{1}{2 c_V^5 c_{a}}+ \frac{2}{3c_{a}(2-c_{a})c_S^5},\quad  {\zeta}_1 \equiv \frac{1}{9c_{a}
  c_S^5(2-c_{a})}\,, 
\\ 
\label{E-func}
{\zeta}_2 &\equiv \frac{4}{3 c_S^3
  c_{a}(2-c_{a})}+\frac{4}{3 c_{a} c_V^3},\quad {\zeta}_3\equiv \frac{1}{6 c_V^5 c_{a}}\,,\\
\label{Z-def}
Z &\equiv \frac{(\alpha_1-2\alpha_2)(1-c_{\sigma})}{3(2c_{\sigma}-c_{a})}\,,\quad  {\cal S}\equiv \frac{m_B s_A+m_A s_B}{m_A+m_B}\,.
\end{align}
Note that the dipole flux is proportional to  ${\zeta}_2$ and to $(s_1-s_2)^2$ (just like in scalar-tensor theories of the Fierz-Jordan-Brans-Dicke type~\cite{1975ApJ...196L..59E,0264-9381-9-9-015,Will:1989sk}). Therefore, it may dominate over GR's quadrupole emission at low frequencies, depending on the sensitivities and 
the coupling parameters of the theory~\cite{PhysRevD.100.083012}.

\section{Solutions for slowly moving stars}\label{sec:exp_V}

In order to compute the sensitivities, we start from the observation that
the metric and \ae ther solutions for a single point particle [Eqs.~\eqref{g00PN}--\eqref{ui_1E} with $\tilde{m}_2=0$]
depend on the sensitivity $\sigma$ already at linear order in the particle's velocity.
Moreover, $\sigma$
regulates the decay of the metric and \ae ther components at large radii and enters already at ${\cal O}(1/r)$. The sensitivity is of course a free parameter in the metric and \ae ther solutions for a point particle, but it can be determined by replacing the point particle with a body of finite size. Once a fully non-linear solution for such a body (e.g., in our case, a NS) has been obtained, one can extract its sensitivity from the asymptotic fall-off of the metric and \ae ther fields. Obviously, since $\sigma$ appears at linear order in velocity, the NS must be moving relative to the preferred foliation. Here, we follow \cite{Yagi:2013ava}
and consider a star in slow motion with respect to the \ae ther,  solve the field equations through linear order in the star's velocity, and extract the sensitivities from the asymptotic decay of the fields.

\subsection{Metric Ansatz}

 Here, we consider the case of a non-spinning NS at rest with a background \ae ther field moving relative to it. The system is in a stationary regime, i.e, there is no dependence of the metric and the \ae ther field on the time coordinate.
For this configuration, letting $v^{i}$ be the velocity of the star relative to the \ae ther, we consider the following ansatz for the metric and the \ae ther 
\begin{align}
  \nonumber  ds^{2} &= e^{\nu(r)}dt^{2} - \left(1- \frac{2 M(r)}{r}\right)^{-1} dr^{2} - r^{2}(d\th^{2} + \sin^{2}\th d\phi^{2})\\
  \label{metric}  &+ 2vV(r,\th)dtdr + 2vS(r,\th)dt d\th + \mathcal{O}(v^{2}),\\ 
  \label{aetherfield}
    U_{\mu} &= e^{\nu(r)/2}\delta^{t}_{\mu} + v W(r,\th)\delta^{r}_{\mu} + v Q(r,\th)\delta^{\th}_{\mu} + \mathcal{O}(v^{2}),
\end{align}
and we will set $G_N=1$  from here on. Note that $M(r)$ has dimensions of length and $M(r)\rightarrow M_{\star}$ as $r$ $\rightarrow \infty$, with $M_\star$ thus being the measured mass of the star.
Here we have adopted a coordinate system that is comoving with the fluid elements of the NS, by aligning the time coordinate vector to the fluid 4-velocity $u^{\mu}$~\cite{Yagi:2013ava}. In these comoving coordinates, the fluid elements are at rest while the \ae ther is moving. The fluid 4-velocity field is 
\be\label{4velocity}
    u^{\mu} = e^{-\nu/2}\delta^{\mu}_{t} .
\ee
 The ansatz of Eqs.~\eqref{metric}--\eqref{4velocity} depends on $M(r)$ and $\nu(r)$ at $\mathcal{O}(v^{0})$ and on four potentials $V(r,\th)$, $S(r,\th)$, $W(r,\th)$ and $Q(r,\th)$ at $\mathcal{O}(v)$. However, one can perform a coordinate transformation of the form
\be
    t' = t + v H(r,\th),
\ee
which allows for any one  of the four potentials to be set to zero while keeping the ansatz valid at $\mathcal{O}(v)$. Here we choose to set $Q = 0$ without loss of generality.

\subsection{Zeroth order in velocity}
From Eqs.~\eqref{aether_ein_eq}--\eqref{aethereqn} let us derive the field equations at zeroth order in velocity, which we will solve to construct the background NS solution. The $(t,t)$, $(r,r)$ and $(\th,\th)$ components of the field equations are the only non-trivial ones and give three independent equations~\cite{Yagi:2013ava} 
\begin{align}
    &16 \frac{dM}{dr} - 4 c_{a}r(r-2 M)\frac{d^{2}\nu}{dr^{2}} - c_{a}r(r-2M)\left(\frac{d\nu}{dr}\right)^{2}\nonumber \\
    &~~~~~~~~~~~~~~~~~~~~~+ 4c_{a}\left(r\frac{dM}{dr} - 2r +3M\right)\frac{d\nu}{dr} = 64 \pi \rho r^{2} \,,\label{eq:v0_eqn1}
    \\
    &c_{a}r^{2}(r-2M)\left(\frac{d\nu}{dr}\right)^{2} + 8r(r-2M)\frac{d\nu}{dr} - 16M = 64\pi r^{3}P \,,\label{eq:v0_eqn2}
    \\
    &\nonumber 4r^{2}(r - 2M)\frac{d^{2}\nu}{dr^{2}} - (c_{a} - 2)r^{2}(r-2M)\left( \frac{d\nu}{dr}\right)^{2} \\
    &~~~~~ - 4r\left(r\frac{dM}{dr} - r + M\right)\frac{d\nu}{dr}- 8r\frac{dM}{dr} + 8M = 64\pi r^{3}P \,,
    \label{eq:v0_eqn3}
\end{align}
where $P(r)$ and $\rho(r)$ are rescaled NS pressure and density respectively (rescaled because $G_N = \frac{2 G}{2-c_a}$). These can be expressed as
\begin{align}
    P \equiv \frac{2 - c_{a}}{2}\tilde{P}, \quad  \rho \equiv \frac{2 - c_{a}}{2} \tilde{\rho} ,
\end{align}
with $\tilde P$ and $\tilde \rho$ representing the pressure and energy density that enter directly in the stress-energy tensor for the matter field, which we take to be of a perfect fluid form 
\begin{align}
T_{\mu \nu}^{\tiny{\mbox{mat}}} = \left(\tilde{\rho} + \tilde{P}\right)u_{\mu}u_{\nu} - \tilde{P}g_{\mu\nu} + \mathcal{O}(v^2)\,. 
\label{eq:T_mat}
\end{align}
Note, that there is a bijective correspondence between the original parametrization $(c_a,c_\theta,c_{\omega},c_\sigma)$ and $(\a_{1},\a_{2},c_{\omega},c_\sigma)$ that can be derived from Eqs.~\eqref{eq:alpha1} and~\eqref{eq:alpha2}.
As discussed in Sec.~\ref{sec:AE1},
 $c_{\sigma} \ll 10^{-15}$ from  gravitational wave observations; we thus rewrite Eqs.~\eqref{eq:v0_eqn1}--\eqref{eq:v0_eqn3} in terms of ($\a_1$, $\a_2$, $c_{\sigma}$, $c_{\omega}$).
 This is justified because the bound on $c_\sigma$ is much stronger than those on the other parameters, which
 are constrained by solar-system and stability requirements (absence of gradient instabilities and vacuum Cherenkov radiation, and positive energy) to satisfy
 \begin{equation}
\alpha_1 < 8 \alpha_2 <  0\,, \quad c_\omega > -\frac{\alpha_1}{2}\,,
\end{equation}
with
$|\alpha_1|\lesssim 10^{-4}$ and $|\alpha_2|\lesssim10^{-7}$,
in the limit $c_\sigma\to 0$.
Upon simplification, we get the modified Tolman-Oppenheimer-Volkoff (TOV) equations
\begin{align}\label{dmdr}
    \nonumber \frac{dM}{dr}&=  \frac{1}{\a_{1}(\a_{1} + 8)r}\left\{ -4\sqrt{r- 2M}(\a_{1} + 8) \sqrt{(-\a_{1}+8)M - 4P\a_{1}\pi r^{3} + 4r} \right.\\
    & \left.-\left( \a_1^{2} + 24\a_{1} +128 \right) M - 16r\left(\a_{1} \pi r^{2} (\a_{1}+2)P - 2\pi r^{2}\a_{1}\rho - \frac{\a_{1}}{2} - 4 \right) \right\} \,,
    \\
    \label{dnudr}
    \frac{d\nu}{dr}&= \frac{1}{\a_{1}(r-2M)r}\left[ -8 \sqrt{r-2M}\sqrt{(-\a_{1} - 8)M - 4P\a_{1} \pi r^{3} + 4r} + 16r -32M\right] \,,
    \\
    \label{dpdr}
    \frac{dp}{dr}&= \frac{1}{\a_{1}(r-2M)r} 4(P + \rho) \left[ \sqrt{r-2M}\sqrt{(-\a_{1} - 8)M - 4P\a_{1} \pi r^{3} + 4r} - 2r + 4M\right]\,.
\end{align}
Modifications to the GR TOV equations can be singled out by expanding the above equations~\eqref{eq:v0_eqn1}--\eqref{eq:v0_eqn3} in a small coupling approximation, i.e., $c_{a} \ll 1$ or $\alpha_1 \ll 1$~\cite{tolman:1939jz,Oppenheimer,Yagi:2013ava}.

\subsection{First order in velocity}
We derive field equations at first order in velocity from Eqs.~\eqref{aether_ein_eq} and~\eqref{aethereqn}, which include the potentials as functions of $r$ and $\th$, at first order in velocity. We can separate variables in $r$ and $\th$ using a Legendre decomposition~\cite{Yagi:2013ava} to obtain
\begin{align}
    V(r,\th) &= \sum_{n} K_{n}(r) P_{n}(\cos\th), \\
    S(r,\th) &= \sum_{n} S_{n}(r) \frac{P_{n}(\cos\th)}{d\th}, \label{eq:Sn} \\
    W(r,\th) &= \sum_{n} W_{n}(r) P_{n}(\cos\th),
\end{align}
where $P_{n}$ is the Legendre polynomial of order $n$. More details on tensor harmonic decomposition can be found in~\cite{Thorne}. By separation of variables, we arrive at $\mathcal{O}(v)$ equations, where only the $(t,r)$ and $(t,\th)$ components of the modified Einstein equations and the $r$ and $\th$ components of the \ae ther field equations are non-trivial. We are only interested in $n=1$ component of Legendre decomposition since these functions determine sensitivities and consequently the change in orbital period. 

Since $c_{\sigma} \ll 10^{-15}$ (c.f. Sec.~\ref{sec:AE1}), we proceed with calculations in the limit $c_{\sigma} \rightarrow 0 $ to obtain~\cite{Yagi:2013ava} 
\allowdisplaybreaks
\begin{align}
    \nonumber \frac{dS_1}{dr} &= \frac{1}{\a_{1}r(r-2M)}\left\{-2 S_1\sqrt{r-2M}(\a_{1}+4)\sqrt{(-\a_{1}-8)M - 4 P \a_{1}\pi r^{3} +4r} - \right.\\
    &\left. (r-2M)\left( J_1e^{\nu/2}c_{\omega}\a_{1} - (3 \a_{1} + 16)S_1 + K_1\a_{1}(c_{\omega}-1)\right)\right\}\,, \label{dsdr}
    \\
    \nonumber \frac{dK_1}{dr} &= \frac{1}{c_{\omega}(r-2M)(\a_1^2 + (2-2\a_{2})\a_{1} - 16\a_{2}) \a_1(\a_1 + 8) r} \left\{ 2 \left[((c_{\omega} + 1)\a_{1} + 2 c_{\omega})J_{1}e^{\nu/2} \right. \right.\\
    \nonumber & \left. - (6\a_{1} + 32)S_1 + c_{\omega}K_1 \a_{1}\right] (\a_1^2 + (2-2\a_{2})\a_{1} - 16\a_{2})(\a_{1}+8)\sqrt{r-2M}\\
    \nonumber &\times \sqrt{(-\a_{1}-8)M - 4P\a_{1}\pi r^{3} + 4r } + \left[-(\a_{1} +8)(r-2M)\a_{1}r \right. \\
    \nonumber &((c_{\omega}+1)\a_1^{2} - 2c_{\omega}(\a_{2}-1)\a_{1} - 16\a_{2}c_{\omega})\left(\frac{\d J_{1}}{\d r}\right) -16\left(-\frac{1}{8}\left(3\left(\left(c_{\omega} + \frac{5}{3}\right)\a_1^{3}  \right. \right. \right. \\
    \nonumber &+\left(\left(-2\a_{2}+\frac{14}{3}\right)c_{\omega}  -\frac{8\a_{2}}{3}+ \frac{8}{3}\right)\a_1^{2} + \left(\left(-\frac{64\a_{2}}{3} + \frac{16}{3}\right)c_{\omega} - \frac{64\a_{2}}{3}\right)\a_{1} \\
    \nonumber &\left. \left.- \frac{128 \a_{2} c_{\omega}}{3} \right) (\a_{1} + 8)M\right) + \left( \pi(\a_{1}+2)\a_{1}r^{2}((c_{\omega}+1)\a_1^{2} - 2c_{\omega}(\a_{2}-1)\a_{1}  \right.\\
    \nonumber &- 16\a_{2}c_{\omega})P - 2\pi \a_{1}r^{2}((c_{\omega}+1)\a_1^{2} - 2c_{\omega}(\a_{2}-1)\a_{1} - 16\a_{2}c_{\omega})\rho\\
    \nonumber& + \frac{1}{4}\left((\a_{1}+8) \left(\left(c_{\omega}+\frac{3}{2}\right)\a_1^{3} + ((-2\a_{2}+4)c_{\omega} - 2\a_{2}+2)\a_1^{2} \right. \right.\\
    \nonumber &\left. \left. \left. \left. \left. + ((-20\a_{2}+4)c_{\omega}-16\a_{2})\a_{1} - 32\a_{2}c_{\omega} \right) \right) \right) r \right) J_1\right] e^{\nu/2} \\
    \nonumber & + 6\left(-\frac{2(\a_{1}+8)^{2}S_1}{3} + c_{\omega}K_1\a_{1}\right) (\a_{1}+8)(\a_1^{2} + (2-2\a_{2})\a_{1} - 16\a_{2})M \\
    \nonumber &-16\left[\left(\pi (\a_1^{2} + (2-2\a_{2})\a_{1}-16\a_{2})(\a_{1}+4) \a_{1}r^{2}P - \frac{11 \a_{2}^{3}}{8}  \right. \right.\\
    &\nonumber \left. +(3\a_{2}-11)\a_1^2 + (40\a_{2}-16)\a_{1} + 128\a_{2}\right)(\a_{1}+8)S_1 + c_{\omega}K_1(\a_1^{2}\\
    &\left. \left.+(2-2\a_{2})\a_{1} - 16\a_{2})\a_{1}\left(\pi r^{2}(\a_{1}+2)P - 2\pi r^{2}\rho + \a_{1}/4 +2 \right)\right]r\right\}\,, \label{dkdr}
    \\
    \nonumber \frac{d^{2}J_1}{dr^2} &= \frac{1}{\sqrt{(-\a_{1}-8)M - 4P \a_{1} \pi r^{3} + 4r} (r -2M)^{3/2} \a_1^{3}r^{2}(\a_{1}+8)}\left(4\left\{ \left[(\a_{1}+8) \right. \right. \right.\\
    \nonumber &\left((\a_1^{2} - 4\a_{1}\a_{2} -32 \a_{2})S_1 + \left(-\frac{\a_1^{2}}{2} + c_{\omega}(\a_{2}-1)\a_{1} + 8\a_{2}c_{\omega}\right)K_1\right)\a_{1} r e^{-\nu/2} \\
    \nonumber & -12\a_1^{2}\left( \left( \frac{1}{24}\a_1^{2} + \a_{1} + \frac{16}{3}\right)M + \left(\a_{1} \pi r^{2}(\a_{1}+2)P - 2 \pi r^{2}\a_{1}\rho + \frac{\a_1^{2}}{24} -\frac{8}{3}\right)r\right)  \\
    \nonumber & \times r \left(\frac{\d J_1}{\d r}\right) + \left(8\left(\frac{\d \rho}{\d r}\right) \a_1^{3} \pi r^{4} + \frac{1}{2}\left((\a_{1} +8)\left( \a_1^{3} + (-8 \a_{2} +56)\a_1^{2} \right. \right. \right. \\
    \nonumber &+\left. \left. (-192 \a_{2} + 128)\a_{1} - 1024 \a_{2}\right) M \right)+ \left( 8 \pi(\a_1^{3} + (-5 \a_{2} + 14)\a_1^{2}  \right.\\
    \nonumber &+ (-56 \a_{2} + 24)\a_{1} -128 \a_{2})\a_{1}r^{2} P + 4 \pi \a_1^{2}r^{2}\left(\a_1^{2} + (-2 \a_{2} + 16)\a_{1} - 16\a_{2} \right. \\
    \nonumber &+ \left. 16 \right)\rho + \left( \frac{\a_1^{3}}{2} + (\a_{2} c_{\omega} - c_{\omega} - 12)\a_1^{2} + (-32 + (8 c_{\omega} + 32)\a_{2})\a_{1} + 256 \a_{2} \right) \\ 
    \nonumber &\times \left. \left.(\a_1 + 8)r \right)J_1\right] \sqrt{r- 2M} \sqrt{(-\a_{1} -8)M - 4P\a_{1} \pi r^{3} + 4r} \\
    \nonumber & + 64\left( \left( \frac{\a_{1}}{4} + 2\right)M + P \a_{1}\pi r^{3} - r\right) \\
    \nonumber &\times \left[ \frac{1}{8} \left(-S_1 \a_{1} \a_{2} r (\a_{1}+8)^{2} e^{-\nu/2} + \a_1^{2} r (\a_{1} + 8)(r - 2M) \frac{\d J_1}{\d r}\right) + J_1\left( \frac{3(\a_{1} +8)}{4} \right. \right. \\
    \nonumber &\times \left( \a_1^{2} + \left(\frac{-8\a_{2}}{3} + \frac{8}{3}\right) \a_{1} - \frac{64 \a_{2}}{3}\right)M +  \left(r^{2}\pi \a_1^{2}(\a_{1} +2)P  \right.\\
    & +\left. \left. \left. \left. \left. \a_1^{2}\pi r^{2}(\a_{1}+2)\rho - \frac{3 \a_1^3}{8}+ (\a_{2} -4)\a_1^{2} + (16 \a_{2} - 8)\a_{1} + 64\a_{2}\right) r \right)\right]\right\}\right),\label{d2jdr2}
\end{align}
where we have defined $J_{n} = W_{n} + e^{-\nu/2}K_{n}$~\cite{Yagi:2013ava}.  With the above set of equations at hand, the next section describes the methods of solving these equations at each order in velocity.

\section{The calculation of the sensitivities}\label{sec:sensit}

 The sensitivities are calculated by solving the coupled differential equations in Eqs.~\eqref{dmdr}-\eqref{dpdr} and Eqs.~\eqref{dsdr}-\eqref{d2jdr2}, which are obtained from the modified Einstein and the \ae ther field equations in a $v\ll1$ expansion at $\mathcal{O}(v^{0})$ and $\mathcal{O}(v)$ respectively~\cite{Yagi:2013ava}. In Secs.~\ref{sub:method1} and \ref{sub:method2} we describe and apply two methods to solve these equations and find the NS sensitivities. The first method, outlined in Sec.~\ref{sub:method1}, was used previously in Ref.~\cite{Yagi:2013ava}, but we will explain how it leads to unstable solutions in particular regions of parameter space. A second method outlined in Sec.~\ref{sub:method2} provides stable results in all regions of parameter space.
 
The $\mathcal{O}(v^0)$ solutions are common to both methods, as they both involve solving $\mathcal{O}(v^0)$ differential equations~\eqref{dmdr}--\eqref{dpdr} numerically once in the interior and then in the exterior of the NS. The initial and boundary conditions to it are obtained by imposing regularity  at the NS center, while imposing asymptotic flatness at spatial infinity respectively. The differential equations are solved from a core radius (i.e. some small initial radius)  to the stellar surface radius, where the pressure goes to zero. These numerical solutions at the NS surface are now used as initial conditions to solve the exterior evolution equations from the stellar surface to an extraction radius $r_b$. Using continuity and differentiability of the solutions, the asymptotic solutions at spatial infinity are matched to the numerical solutions evaluated at $r_b$. This gives the observed mass of the NS and the integration constant corresponding to $\nu(0)$ (obtained by solving the $\mathcal{O}(v^0)$ differential equations~\cite{Yagi:2013ava}) which will be used in solving the $\mathcal{O}(v)$ equations discussed further.    

\subsection{Method 1: Direct Numerical Solutions}\label{sub:method1}

In this method, the aforementioned $\mathcal{O}(v)$ differential equations are solved in two regions, the interior of the star, and the exterior. The initial conditions at $\mathcal{O}(v)$ are obtained by solving the corresponding differential equations asymptotically about a core radius, while imposing regularity at the core, and asymptotically about spatial infinity, while imposing asymptotic flatness~\cite{Yagi:2013ava}. In both cases, the solutions depend on integration 
constants -- $\tilde{C}$ and $\tilde{D}$ in the interior asymptotic solution and $\tilde{A}$ and $\tilde{B}$ in the exterior asymptotic solution -- that must be chosen so as to guarantee that the numerical interior and exterior solutions are continuous and differentiable at the stellar surface, where pressure becomes significantly smaller than their core values.    

As defined above, the global solution reduces to finding the right constants $(\tilde{A},\tilde{B},\tilde{C},\tilde{D})$, which in turn is a shooting problem. In practice, Ref.~\cite{Yagi:2013ava} solved this shooting problem by first picking two sets of values for interior constants, $\vec{c}_{(1)} = (\tilde{C}_{(1)},\tilde{D}_{(1)})$ and $\vec{c}_{(2)} = (\tilde{C}_{(2)},\tilde{D}_{(2)})$, and then solving the interior equations twice from the core radius $r_{c}$ to the NS surface $R_{\star}$ to find the solutions $\vec{f}_{(1)}^{\interior}(r) = [S_1^{(1,\interior)}(r),K_1^{(1,\interior)}(r),J_1^{(1,\interior)},J_1'^{(1,\interior)}(r)]$ and $\vec{f}_2^{\interior}(r) = [S_1^{(2,\interior)}(r),K_1^{(2,\interior)}(r),J_1^{(2,\interior)}(r),J_1'^{(2,\interior)}]$. Then, each interior numerical solution is evaluated at the stellar surface and used as initial conditions for a numerical evolution in the exterior, leading to two exterior solutions $\vec{f}_1^{\ext}(r) = [S_1^{(1,\ext)}(r),K_1^{(1,\ext)}(r),J_1^{(1,\ext)}(r),J_1'^{(1,\ext)}(r)]$ and $\vec{f}_2^{\ext}(r) = [S_1^{(2,\ext)}(r),K_1^{(2,\ext)}(r),J_1^{(2,\ext)}(r),J_1'^{(2,\ext)}(r)]$. 

The global solutions $\vec{f}_{1,2}^{\glo}(r) = \vec{f}_{1,2}^{\interior}(r) \cup \vec{f}_{1,2}^{\ext}(r)$ are then automatically continuous and differentiable at the surface, but in general they will not satisfy the boundary conditions at spatial infinity. Because of the linear and homogeneous structure of the differential system, one can find the correct global solution through linear superposition
\begin{equation}\label{linearsuper}
    \vec{f}^{\glo}(r; C',D') = C' \vec{f}^{\glo}_{1}(r) + D' \vec{f}^{\glo}_{2}(r)\,,
\end{equation}
where $C'$ and $D'$ are new constants, chosen to guarantee that $\vec{f}^{\glo}$ satisfies the correct asymptotic conditions near spatial infinity, which in turn depend on $(\tilde{A},\tilde{B})$, i.e.
\begin{equation}
    \vec{f}^{\glo}(r_b; C',D') = \vec{f}^{\glo,\infty}(r_b; \tilde A, \tilde B)\,,
\end{equation}
where $r_b \gg R_{\star}$ is the matching radius, $\vec{f}^{\glo}(r_b; C',D')$ is given by Eq.~\eqref{linearsuper} evaluated at $r=r_b$ (which depends on $(C',D')$) and $\vec{f}^{\glo,\infty}(r_b; \tilde A, \tilde B)$ is the asymptotic solution to the differential equations near spatial infinity evaluated at the matching radius (which depends on $(\tilde{A},\tilde{B})$).  

With this at hand, one can calculate the NS sensitivities via~\cite{Yagi:2013ava}
\begin{equation}\label{sens_method1}
    \sigma = 2 \tilde{A} \frac{\a_{1}}{\a_{1}+8} \,,
\end{equation}
where $\tilde{A}$ is the coefficient of $1/r$ in the near-spatial infinity asymptotic solution of $W_1^{\ext}$ such that
\begin{equation}
    W_1^{\ext}(r) = \tilde{A}\frac{M_{\star}}{r} +  \mathcal{O}\left(\frac{M_{\star}^2}{r^2}\right)\,,
\end{equation}
while we recall that   $\a_{1} \lesssim   10^{-4}$ (c.f.  Sec.~\ref{sec:AE1}). Because of the latter constraint, it is obvious that the sensitivities are essentially controlled by $\sigma \approx \tilde{A}\alpha_1$, so the numerical stability of its calculation relies entirely on the numerical stability of the calculation of this coefficient. Unfortunately, as we show below, this calculation is not numerically stable in the region of parameter space we are interested in. 

\begin{figure}
\centering
\includegraphics[scale=0.45]{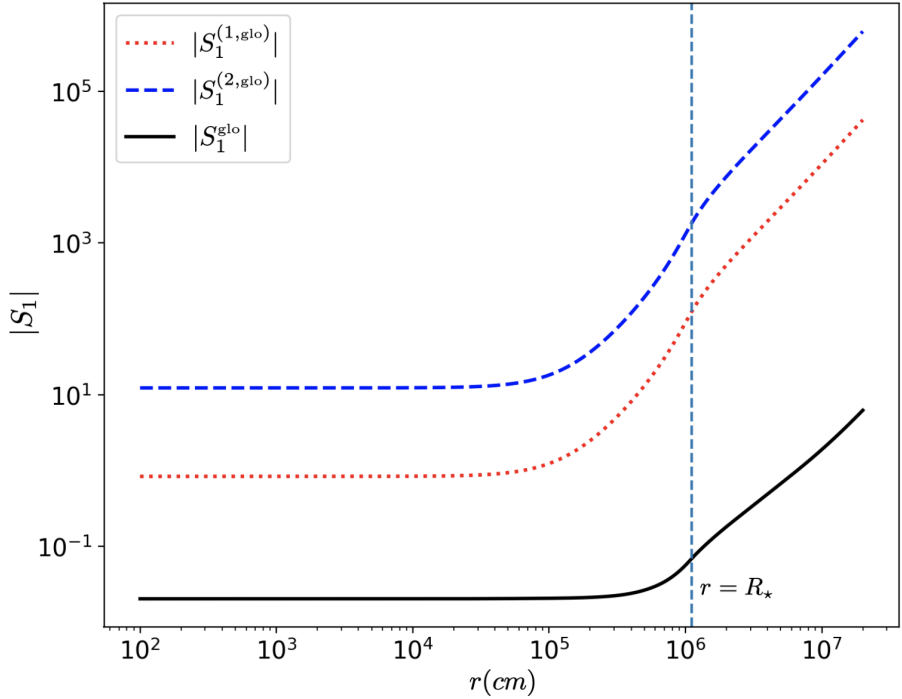}
\caption{ The metric function |$S_1$| is plotted against the radius in the entire numerical domain, where the radius of the star is at 11.1 km (vertical dashed line). Observe that both of the trial solutions $S_1^{(1,\glo)}$ and $S_1^{(2,\glo)}$ diverge at spatial infinity. Hence, the global linearly combined solution $S_1^{(\glo)}$ shows a diverging behaviour representing numerical instabilities in the calculation of sensitivities.}
\label{fig:method1}
\end{figure}

Figure \ref{fig:method1} shows $S_1$ 
as a function of radius, assuming $(\a_{1},\a_2,c_{\omega}) = (10^{-4}, 4\times 10^{-7}, -0.1)$, and setting $(r_c,r_b) = (10^2, 2 \times 10^7)$ cm. Observe that both $S_1^{(1,\glo)}$ and $S_1^{(2,\glo)}$ diverge at spatial infinity, so in order to find an $S_1^{\glo}$ that is finite at spatial infinity, a very delicate cancellation of large numbers needs to take place. This cancellation needs to lead to $\tilde{A}\alpha_1 \approx 0$ but in general $\tilde{A}\alpha_1 \neq 0$, since $\sigma \approx \tilde{A}\alpha_1/4 \ll 1 \neq 0$, and precisely by how much $\tilde{A}\alpha_1$ deviates from $0$ is what determines the value of the sensitivity. We find in practice that $\sigma$ is highly sensitive to the accuracy of the numerical algorithm used to solve for $\vec{f}_{(1,2)}^{\glo}$, as well as the choice of $r_c$, $r_b$ and the value of $p(R_{\star})$ that defines the stellar surface. Figure~\ref{fig:method1} is in the parameter region that is outside of interest but it indicates how sensitive the calculations are to the aforementioned cancellation,  making it difficult to find numerically stable solutions.

\subsection{Method 2: Post-Minkowskian Approach}\label{sub:method2}
Given that the first method does not allow us to robustly compute the sensitivities in the regime of interest, we developed a new post-Minkowskian method, which we describe here. In this method, the background $O(v^0)$ equations are solved by direct integration, as done in method 1. The differential equations at $O(v)$, however, are expanded in compactness $\mathcal{C}$ and solved order by order. This is a post-Minkowskian approximation because the compactness always appears multiplied by $G/c^2$, so in this sense it is a weak-field expansion. NSs are not weak-field objects, but their compactness is always smaller than $\sim 1/3$ (and usually between $[0.1,0.3]$), so provided enough terms are kept in the series, this approximation has the potential to be valid. Moreover, and perhaps more importantly, we will show below that such a perturbative scheme stabilizes the numerical solution for the NS sensitivities. 

The procedure presented above is not technically a standard post-Minkowskian series solution because the background equations (or their solutions) are not expanded in powers of $\mathcal{C}$. Had we expanded in powers of $\mathcal{C}$ everywhere, 
we would have encountered terms in the differential equations with derivatives of the equation of state (EoS). Such derivatives would introduce numerical noise because ``realistic'' (tabulated) EoSs are not usually smooth functions, potentially introducing steep jumps, see, e.g.,~\cite{Yagi:2013mbt}. By not expanding the $\mathcal{O}(v^{0})$ differential equations, we are implicitly adding higher order terms in compactness, so this procedure could be seen as a resummation technique. 

Through this approach, the differential equations at $\mathcal{O}(v^{1})$ turn into $n$ sets of differential equations for an expansion carried out to ${\cal{O}}(\mathcal{C}^n)$, with $n$ therefore labeling the compactness order. In order to derive these equations, however, one must first establish the order of the background solutions, which can be shown to satisfy 
\begin{align}
    &M(r) = \mathcal{O}(\mathcal{C})\,, \hspace{1.9cm}
    P(r) = \mathcal{O}(\mathcal{C}^2)\,,\\
    &\rho(r) = \mathcal{O}(\mathcal{C})\,, \hspace{0.7cm} \textrm{and} \hspace{0.7cm}
    \nu(r) = \mathcal{O}(\mathcal{C})\,, 
\end{align}
by looking at the differential equations these functions obey at $\mathcal{O}(v^{0})$. The metric perturbation functions at $\mathcal{O}(v^{1})$ are then expanded in powers of compactness through 
\begin{align}
    Y_{i}(r) &= \sum_{j=1}^{\text{n}}Y_{ij}\mathlarger{\epsilon}^j\,, \label{C_expan}
\end{align}
where $Y_{i}$ $\equiv$ ($S_1, K_1, W_1$), $\epsilon$ is a bookkeeping parameter of $\mathcal{O}(\mathcal{C})$ and $j$ indicates the order of $\mathcal{C}$ to be summed over. We work with $W_1(r)$ instead of $J_1(r)$ to avoid introducing numerical error during the conversion between these two functions. 

Using these expansions in the differential equations at $\mathcal{O}(v)$, and re-expanding them in powers of compactness, one finds $n$ sets of differential equations. At $\mathcal{O}(\mathcal{C})$, the differential system becomes
\begin{align}
    \frac{d S_{11}}{d r}&=\frac{2 r K_{11}-2 r\left(S_{11}+\alpha_{1} W_{11}\right)-\left(4+\alpha_{1}\right) M}{2 r^{2}}\,,\label{ds_firstC}
    \\
    \nonumber\frac{dK_{11}}{dr} &= \frac{-1}{2 c_{\omega} r^{2}\a_{1}}\left[ 8 c_{\omega}\pi r^{3}\a_{1}\rho + 4 c_{\omega}r\a_{1}(K_{11} - S_{11}) - 4r\a_1^{2}W_{11} \right.\\
    \nonumber & + c_{\omega}(8\a_{1} + \a_1^{2} - 16\a_{2} - 2\a_{1}\a_{2})M  - 4c_{\omega}r^{2}\a_{1}W_{11}' \\
    &\left.- 2r^{2}(\a_1^{2} + c_{\omega}\a_1^{2} - 16 c_{\omega}\a_{2} - 2c_{\omega}\a_{1}\a_{2})W_{11}'\right]\,, \label{dk_firstC}
    \\
    \frac{d^{2}W_{11}}{dr^2} &= \frac{2 W_{11}}{r^{2}} \,,\label{dw_firstC}
\end{align}
where $S_{11}(r)$, $K_{11}(r)$ and $W_{11}(r)$ are metric functions at $\mathcal{O}(\mathcal{C})$, and recall that the density $\rho(r)$ is related to pressure through the EoS as $\rho(r)$ = $\rho(P(r))$. We note that $W_{11}(r)$ is decoupled from $S_{11}(r)$ and $K_{11}(r)$, and so we can solve Eq. (\ref{dw_firstC}) separately and analytically in the regions $r \leq r_b$ and $r \geq r_b$.
The solutions to them are $W_{11}(r) = \tilde{D}_{1} r^2$ for $r \leq r_b$ and $W_{11}^\infty(r) = \tilde{A}_{1}/r$ for $r \geq r_b$, where $\tilde{A}_1$ and $\tilde{D}_1$ are integration constants.
By requiring continuity and differentiability of metric functions $W_{11}(r)$ and $W_{11}'(r)$, we match the solutions at the extraction radius $r_b$. This fixes the values of two integration constants $\tilde{A}_{1} = 0$ and $\tilde{D}_{1} = 0$.  

The remaining two equations, namely Eqs.~(\ref{ds_firstC}) and (\ref{dk_firstC}),  are solved numerically with initial conditions obtained using regularity at the NS center and asymptotic flatness at spatial infinity. At the center, we have
\begin{align}
    \nonumber S_{11}(r) &= \tilde{C}_{1} - \frac{1}{120 \a_{1}}\pi \left[ (240\a_{1} + 40\a_1^{2} -128 \a_{2} -16\a_{1}\a_{2} )\rho_{c} \right. \\
    \nonumber & \left.+ (48\a_{1}\a_{2} -72 \a_1^{2} -15 \a_1^{3} + 6\a_1^{2}\a_{2})p_{c}\right]r^{2} \\
    \nonumber & +\frac{1}{1260}\pi^{2}\left[(-160\a_{1} - 35\a_1^{2} + 80\a_{2} + 10\a_{1}\a_{2})\rho_{c}^{2} \right.\\
    \nonumber &+ (288\a_{2} -576 \a_{1} -126 \a_1^{2} + 36\a_{1}\a_{2})p_{c}^{2}  \\
    &\left. + (336\a_{2} - 672\a_{1}  -147\a_1^{2} + 42\a_{1}\a_{2})\rho_{c}p_{c}\right]r^{4} + \mathcal{O}(r^6)  \,, \label{sint_firstC}
    \\
    \nonumber K_{11}(r) &= \tilde{C}_{1} - \frac{1}{120 \a_{1}}\pi \left[ (400\a_{1} + 40\a_1^{2} -384 \a_{2} -48\a_{1}\a_{2} )\rho_{c}\right.\\
    \nonumber & \left.+ (144\a_{1}\a_{2} -96 \a_1^{2} -15 \a_1^{3} + 18\a_1^{2}\a_{2})p_{c}\right]r^{2} \\
    \nonumber & + \frac{1}{1260}\pi^{2}\left[(400\a_{2} -240\a_{1} -35\a_1^{2} + 50 \a_{1}\a_{2})\rho_{c}^{2} \right.\\
    \nonumber &+ (1440\a_{2} -864\a_{1} -126\a_1^{2} + 180\a_{1}\a_{2})p_{c}^{2} \\
    &\left.  + (1680 \a_{2} -1088\a_{1} - 147\a_1^{2}  +210 \a_{1}\a_{2})\rho_{c}p_{c}\right]r^{4} + \mathcal{O}(r^6) \,,\label{kint_firstC}
\end{align}
where $\tilde{C}_1$ is an integration constant\footnote{\label{note1}Equations~\eqref{sint_firstC} and~\eqref{kint_firstC} (and also Eqs.~\eqref{sext_firstC} and~\eqref{kext_firstC}) contain terms higher than $\mathcal{O}(C)$ because the background functions are not expanded in a series of $C$ and thus contain higher order contributions.}. At spatial infinity, we have
\begin{align}
    \nonumber S_{11}^{\infty}(r) &= - \frac{\tilde{B}_{1}}{2r^{3}} -\frac{1}{2 c_{\omega}r\a_{1}}\left[ \tilde{A}_{1}(\a_1^{2} -2 c_{\omega}\a_{1} -2 \a_1^{2}c_{\omega} + 16c_{\omega}\a_{2} + 2c_{\omega}\a_{1}\a_{2}) \right.\\
    \nonumber & \left. + c_{\omega}(8\a_{2} -6\a_{1} -\a_1^{2} + \a_{1}\a_{2})M_{\star}\right]+ (16\a_{2} -8\a_{1} - \a_1^{2} + 2 \a_{1}\a_{2})\frac{M_{\star}^{2}}{64 r^{2}}\\
    \nonumber& + (16\a_{2} -16\a_{1} - 3\a_1^{2} + 2\a_{1}\a_{2})\frac{M_{\star}^{3}}{192r^{3}} \\
    & + (8\a_{2} -2\a_{1} + \a_{1}\a_{2})\ln\left(\frac{r}{M_{\star}}\right)\frac{M_{\star}^{3}}{48 r^{3}} + \mathcal{O}\left(\frac{M_{\star}^4}{r^4}\right)\,, \label{sext_firstC}
    \\
    \nonumber K_{11}^{\infty}(r) &= \frac{\tilde{B}_{1}}{r^{3}} + \frac{4M_{\star} + 2\tilde{A}_{1}\a_{1} + M_{\star}\a_{1}}{2r} - (\a_1^{2} + 16\a_{2} + 2\a_{1}\a_{2})\frac{M_{\star}^2}{64r^{2}} \\
    &+ (2\a_{1} - 8\a_{2}-\a_{1}\a_{2})\ln\left(\frac{r}{M_{\star}}\right)\frac{M_{\star}^{3}}{24 r^{3}} + \mathcal{O}\left(\frac{M_{\star}^4}{r^4}\right)\, \label{kext_firstC}
\end{align}
where $\tilde{B}_1$ is an integration constant and $M_{\star}$ is the mass of the star. 

We next explain how to solve Eqs.~(\ref{ds_firstC}) and (\ref{dk_firstC}) to construct the solution for $S_{11}$ and $K_{11}$. First, homogeneous solutions are given by $S_{11}^{\textrm{hom}} = K_{11}^{\textrm{hom}} = \tilde C_1$. Next, one can construct particular solutions $S_{11}^{\textrm{part}}$ and $K_{11}^{\textrm{part}}$ by setting $\tilde C_1 = 0$ and numerically integrate the equations from $r_{c}$ to $R_{\star}$. 
We then use the numerically calculated interior solutions, evaluated at $R_\star$, as the initial conditions to solve the exterior evolution equations with zero pressure and density from $R_{\star}$ to $r_{b}$. 
The true solutions are simply the sum of the homogeneous and particular solutions, namely
\begin{align}
    S_{11}(r) &= \tilde{C}_{1} + S_{11}^{\textrm{part}} \,,\\
    K_{11}(r) &= \tilde{C}_{1} + K_{11}^{\textrm{part}} \,. 
\end{align}
By requiring continuity and differentiability of all metric functions, we match the true numerical solution to the analytic asymptotic solution in Eqs.~\eqref{sext_firstC} and~\eqref{kext_firstC} at $r_{b}$. Applying this matching condition gives the values of $\tilde{B}_{1}$ and $\tilde{C}_{1}$.

Now let us focus on the solution to $\mathcal{O}(\mathcal{C}^{2})$ differential equations. The equation for the metric function $W_{12}(r)$ is 
\begin{align}
    \nonumber\frac{d^{2}W_{12}}{dr^2} &= \frac{2 W_{12}}{r^{2}} + 4\pi r W_{11}\rho' - 2\pi \rho\left( -2K_{11} + 2S_{11} + (4 +\a_{1} + \frac{2\a_{1}}{c_{\omega}}) W_{11} - 6r W_{11}'\right)\\ 
    \nonumber& - 2\pi \rho(6 + \a_{1})\frac{M}{r} -  W_{11}'\left(4 + \a_{2} + \frac{8\a_{2}}{\a_{1}}\right)\frac{M}{r^{2}} \\
    \nonumber& - \left(3K_{11} - 3S_{11} -10 W_{11} - 2\a_{1}W_{11} - \frac{2\a_{1}W_{11}}{c_{\omega}}\right)\frac{M}{r^{3}}  \\
    & + \left(5 + \a_{1} +\frac{\a_{2}}{2} + \frac{4\a_{2}}{\a_{1}}\right)\frac{M^{2}}{r^{4}} ,\label{d2w_secondC}
\end{align}
where $\rho'(r)$ is the derivative of $\rho(P(r))$ obtained from the EoS. This equation is decoupled from the remaining metric functions at $\mathcal{O}(\mathcal{C}^{2})$, $S_{12}$ and $K_{12}$, and can be solved numerically on its own. The initial condition obtained at the center of the NS is 
\begin{align}
    \nonumber W_{12}(r) &= \tilde{D}_{2}r^{2} + \frac{1}{529200 \a_{1}}\pi^2 r^{4}\left[ 560 \rho_{c}^{3}\pi r^{2}\a_{1}(\a_{1}+8)(5\a_{1} + 4\a_{2}) \right.\\
    \nonumber& - 27 p_{c}^{2}\a_1^{2}(70(8\pi p_{c}r^{2} - 7)\a_1^{2} - 8 \pi p_{c}r^{2}\a_{1}(\a_{2}-422) + 49\a_{1}(\a_{2}-62) \\
    \nonumber&+ 8(49 - 8\pi p_{c}r^{2})\a_{2}) - 252\rho_{c}p_{c}\a_{1}\left\{70p_c \pi r^2 \a_1^3 + \a_1^2(70 - p_{c}\pi r^{2}(\a_{2} - 382)) \right. \\
    \nonumber&\left.+ 64(7 - 4 \pi p_{c}r^{2})\a_{2} - 8\a_{1}(5\pi p_{c}r^{2} (\a_{2}+8) - 7(\a_{2}+10))\right\} \\
    \nonumber&- 12\rho_{c}^{2} \left( 350\pi p_{c}r^{2}\a_1^{4} - 5 \pi p_{c}r^{2}\a_1^{3}(-226 + \a_{2}) - 31360\a_{2}  \right.\\
    &\left. \left.- 784\a_{1}(-40 + (5 + 8\pi p_{c}r^{2})\a_{2}) - 8\a_1^{2}(-490 + \pi p_{c}r^{2}(980 + 103\a_{2}))\right)\right] + \mathcal{O}(r^6),\label{wint_secondC}
\end{align}
where $\tilde{D}_2$ is an integration constant. The boundary condition to Eq.~\eqref{d2w_secondC} at spatial infinity is 
\begin{align}
    \nonumber W_{12}^{\infty}(r) &= \frac{\tilde{A}_{2}}{r} + \frac{1}{320r^{3}\a_{1}c_{\omega}}\left[ \tilde{A}_{1} M_{\star}(40 r - M_{\star}\a_{1})(\a_1^{2} + c_{\omega}(4\a_1^{2}  \right.\\
    \nonumber & + \a_{1}(34 - 4\a_{2}) - 32 \a_{2})) + c_{\omega}(80 M_{\star}^{2}r (\a_{1}+8)(\a_{1}-\a_{2}) \\
    & \left.+ M_{\star}^{3}\a_{1}(-4\a_1^{2} + 56 \a_{2} + \a_{1}(-38 + 7\a_{2})))\right] + \mathcal{O}\left(\frac{M_{\star}^4}{r^4}\right) ,\label{wext_secondC}
\end{align}
where $\tilde{A}_2$ is an integration constant. To construct the solution, we first note that the homogeneous solution is given by $W_{12}^\mathrm{hom} = \tilde{D}_{2} r^{2}$.
Next, we set $\tilde{D}_{2} = 0$ 
and find the particular solution $W_{12}^{\textrm{part}}(r)$ numerically in the interior of the NS by solving Eq.~\eqref{d2w_secondC}. This interior solution evaluated at the NS surface now serves as initial conditions to solve the differential equations in the exterior up to the boundary radius $r_{b}$. The correct solution in the entire numerical domain is then
\begin{equation}
    W_{12}(r) = \tilde{D}_{2}r^{2} + W_{12}^{\textrm{part}}(r),
\end{equation}
where the values of $\tilde{A}_{2}$ and $\tilde{D}_{2}$ are obtained using the matching condition at $r_b$. The equations for $S_{12}$ and $K_{12}$ are solved similar to the way Eqs.~(\ref{ds_firstC}) and (\ref{dk_firstC}) are solved, so we omit a more detailed description here for brevity. We can use the above method to solve differential equations at higher order in $\mathcal{C}$.

\subsection{Comparison between numerical and analytical approaches} 

\subsubsection{Tabulated APR4 EoS}

The sensitivity in the \ae ther theory $\sigma$ for an isolated NS depends on the EoS chosen, and here we perform the calculations of the previous section for the APR4 tabulated EoS~\cite{ap4}. The results are representative of what one finds with other EoSs. 

Eq.~\eqref{sens_method1} gives the expression of sensitivity in terms of the integration constant $\tilde{A}$~\cite{Yagi:2013ava}
where $\tilde{A}$ can be expressed as
\begin{align}
	\tilde{A} &\equiv \frac{1}{M_{\star}}\sum_{j=2}^{\text{n}} \tilde{A}_{j} \mathlarger{\epsilon}^{j},
\end{align}
where $n$ is the order of the compactness expansion, with $\tilde{A}_1=0$. The coefficients $\tilde{A}_{j}$ can be calculated numerically as described in the previous subsection. 
Notice that the leading contribution to $\tilde A$ (and hence to the sensitivities) is of $\mathcal{O}(\epsilon)$, since $M_\star = \mathcal{O}(\epsilon)$.

\begin{figure}
    \centering
    \includegraphics[scale = 0.41]{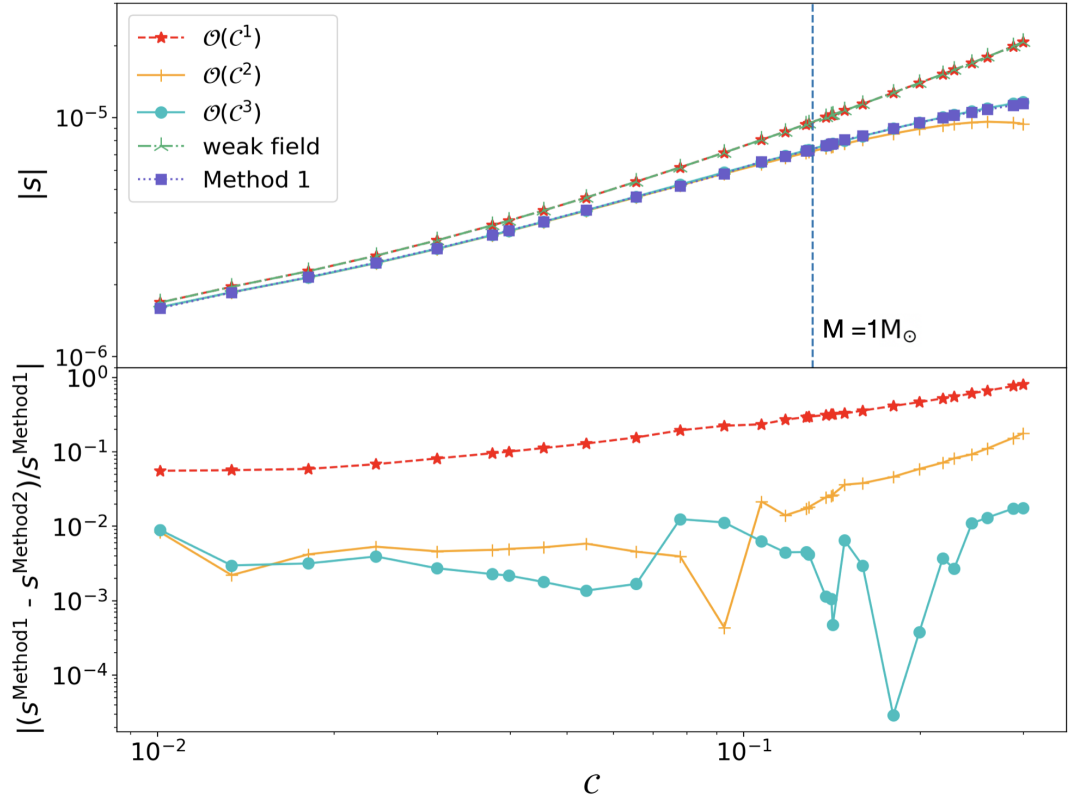}
    \caption{(Top) Sensitivity as a function of compactness using the APR4 EoS, for various post-Minkowskian truncation orders at $(\a_{1},\a_{2},c_{\omega},c_\sigma) = (10^{-4}, 4 \times 10^{-7}, 10^{-4},0)$. At leading order in $\mathcal{C}$, the sensitivity curve overlaps with that computed analytically in the weak field limit in~\cite{Foster:2006az} (Eq.~\eqref{sens_wf}). As the compactness order is increased, the sensitivity curve starts to converge toward the solution found with method 1. (Bottom) Fractional difference between the sensitivities at different order of compactness and those found from method 1. Observe that when $n = 3$, the truncated post-Minkowskian series is already an excellent approximation. The vertical dashed line corresponds to the compactness of a $1M_\odot$ NS.} 
    \label{fig:orderbyorderinC}
\end{figure}

The calculation of the sensitivity as described above requires one to choose the truncation order $n$ of the post-Minkowskian expansion. We will choose $n$ by the sensitivities computed  from methods 1 and 2 in a regime of parameter space where method 1 yields stable results~\cite{Yagi:2013ava}. In particular, we will focus on the choice $(\a_{1},\a_{2},c_{\omega},c_\sigma) = (10^{-4}, 4 \times 10^{-7}, 10^{-4},0)$. Figure~\ref{fig:orderbyorderinC} compares the sensitivities computed with the two methods with this parameter choice. Observe that as the order of post-Minkowskian approximation increases (i.e.~as $n$ increases), the curves approach the method 1 solution, but in an oscillatory manner. The bottom panel of Fig~\ref{fig:orderbyorderinC} shows the stability of post-Minkowskian method at order $n = 3$.\\

In the weak field limit, the sensitivity can be well-approximated as the ratio of the binding energy to the NS mass $(\Omega/M_*)$ through~\cite{foster:2007gr}
\begin{equation}\label{sens_wf}
    s_\mathrm{wf} =  \left(\a_{1} - \frac{2}{3}\a_{2}\right)\frac{\Omega}{M_{\star}} \,,
\end{equation}
where the stellar binding energy $\Omega$ is~\cite{Foster:2006az} 
\begin{equation}
    \Omega = - \frac{1}{2}\int d^{3}x \rho(r) \int d^{3}x'\frac{\rho(r')}{|\mathbf{x - x'}|} \,,
\end{equation}
with $r = |\mathbf{x}|$  and $r' = |\mathbf{x'}|$. We can use a Legendre expansion of the Green's function to evaluate this integral, and to leading order in $\mathcal{C}$, we find a result that is identical to that computed in the weak field limit by~\cite{Foster:2006az}. This can also be seen numerically in Fig.~\ref{fig:orderbyorderinC}, where the weak field curve coincides with $\mathcal{O}(\mathcal{C}^{1})$ post-Minkowskian approximation. 

\

\subsubsection{Tolman VII EoS}

We now focus on the sensitivity of a NS using the Tolman VII EoS. The latter is an analytic model that accurately describes  non-rotating NSs~\cite{tolman:1939jz}
by 
the energy density profile 
\begin{equation}
    \rho(r) = \rho_{c}\left( 1 - \frac{r^2}{R_{\star}^2} \right) \,.
\end{equation}
The advantage of using the Tolman VII EoS is that the background solution is known analytically in GR~\cite{tolman:1939jz,Jiang:2019vmf}.
We expand analytically both the $\mathcal{O}(v^{0})$ and $\mathcal{O}(v)$ equations order by order in compactness. The sensitivity obtained is then
\begin{align}\label{tolman_sens_C}
    \nonumber s &= \frac{5}{21} \mathcal{C} \left(-3\a_{1} + 2\a_{2}\right) \\
    \nonumber& + 5\left(\frac{573\a_1^{3} + \a_1^{2}(67669 - 764 \a_{2})+ 96416\a_{2}^{2} + 68\a_{1}\a_{2}(-2632 + 9\a_{2})}{252252 \a_{1}}\right) \mathcal{C}^{2}\\
    \nonumber &+ \frac{1}{1801079280 c_{\omega} \a_1^{2}} \Big\{(4 \a_{1})^2 (8 + \a_{1}) (36773030 \a_1^2 - 39543679 \a_{1} \a_{2} \\
   	\nonumber& +11403314\a_2^2) + c_{\omega}\left[-1970100 \a_1^5 + 13995878400 \a_{2}^3 \right.\\
   	\nonumber& + 640 \a_1 \a_{2}^2 (-49528371 + 345040\a_2) + 5\a_1^4(-19596941 + 788040\a_{2})\\
   	\nonumber& + \a_1^3(-2699192440 + 440184934\a_2 -5974000 \a_{2}^2)\\
   	&\left. 16\a_1^2 \a_2 (1294533212 - 29152855\a_2 +212350\a_2^2 )\right]\Big\}\mathcal{C}^{3} + \mathcal{O}(\mathcal{C}^4)\,.
\end{align}
Note that the above expression is not regular in the limit of $\a_1 \to 0$ while keeping $\a_2$ finite or $c_\omega \to 0$ while keeping $\a_1$ or $\a_2$ finite. This is a known feature of Einstein-\ae ther theory, which recovers GR only when a certain combination of coupling constants is taken to zero at a specific rate. 

With this EoS, the compactness can be expressed as a function on $\Omega/M_{\star}$ as 
\begin{equation}
\mathcal{C}= -\frac{7 \Omega}{5 M_\star}+\frac{35819 \a_1 \Omega^3}{85800
   M_\star^3} + \mathcal{O}\left(  
  \frac{\Omega^4}{M_\star^4} \right)\,.
\end{equation}
Here $\mathcal{C}$ and $M_{\star}$ are the observed values with \ae ther corrections included.
With this at hand, we can rewrite the sensitivity as a function of $\Omega$ to find
\begin{align}\label{tolman_sens_be}
    \nonumber s &= \frac{(3\a_{1} + 2\a_{2})}{3}\frac{\Omega}{M_{\star}} \\
    \nonumber& + \left(\frac{ 573\a_1^3 + \a_1^2(67669 - 764\a_{2}) + 96416\a_{2}^2 + 68 \a_{1}\a_{2}(9\a_{2} - 2632)}{25740 \a_{1}}\right)\frac{\Omega^2}{M_{\star}^2}\\
    \nonumber& + \frac{1}{656370000 c_{\omega}\a_1^2}\Big\{ -4\a_{1}^2(\a_{1}+8)\left[36773030\a_1^2 - 39543679\a_{1}\a_{2} \right.\\
    \nonumber&\left.+ 11403314\a_{2}^2\right] + c_{\omega}\left[1970100\a_1^5 - 13995878400 \a_{2}^3\right. \\
    \nonumber& - 640 \a_1 \a_{2}^2(-49528371 + 345040 \a_2) - 5\a_1^4(19548109 + 788040\a_2)\\
    \nonumber& - 16 \a_1^2 \a_2 (1294533212 - 29152855 \a_2 + 212350 \a_{2}^2)\\
    & \left.+ \a_1^3(2699192440 - 309701434 \a_2 + 5974000 \a_{2}^2)\right]\Big\}\frac{\Omega^3}{M_{\star}^3} +\mathcal{O}\left(\frac{\Omega^4}{M_\star^4}\right) \,.
\end{align}
Note that this expression matches identically to that of~\cite{foster:2007gr} when working to leading order in the binding energy. 

\begin{figure}[ht!]
    \centering
    \includegraphics[scale = 0.43]{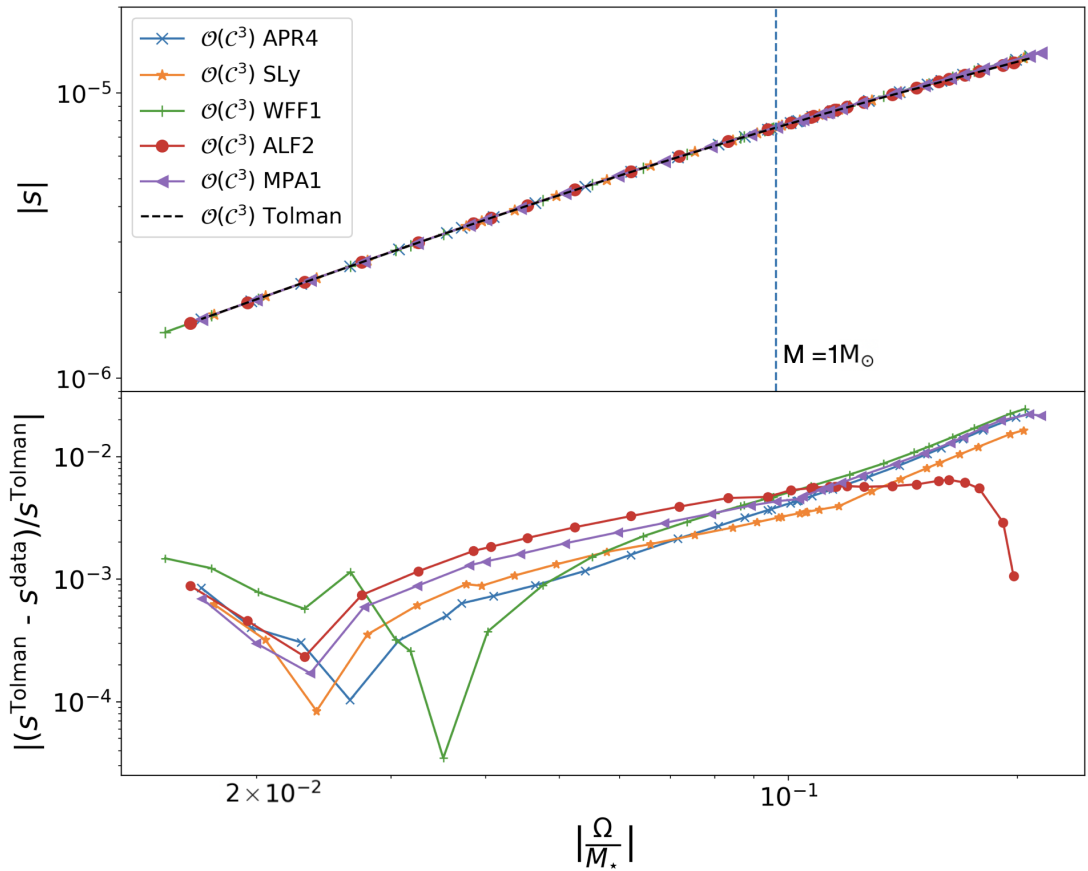}
    \caption{ 
    Top panel shows the plot of sensitivity as a function of binding energy for different EoS including Tolman VII (Eq.~\eqref{tolman_sens_be}) valid to $\mathcal{O}(\mathcal{C}^3)$. The bottom panel shows the relative fractional difference between the EoS from data and the Tolman case, which represents the EoS variation in the relations. Observe that the universality holds to better than 3\%.}
    \label{fig:universal_eos}
\end{figure}

One may wonder whether the above analytic expression is capable of approximating the sensitivity when using other EoSs. Figure~\ref{fig:universal_eos} shows the absolute magnitude of the sensitivity as a function of $\Omega$ computed analytically with Eq.~\eqref{tolman_sens_be}, as well as numerically with six other EoSs. Here we have chosen to work in a different region of parameter space, namely $(\a_{1},\a_{2},c_{\omega}) =(-10^{-4},-4 \times 10^{-7},10^{-3})$, where we obtain a stable smoothly varying sensitivity curve. Observe that the sensitivities differ by less than 3$\%$, exhibiting an approximate universality already discovered in~\cite{Yagi:2013ava} as a function of compactness.
Given these results, in all future calculations we will use the analytic sensitivities computed with the Tolman VII EoS.
 
\section{Constraints from binary pulsars and triple systems} \label{sec:constr}

The majority of millisecond pulsars are found in binary and triple systems. The orbital dynamics of these systems modulate the time of arrival of radio waves and allow for precise measurements of the orbital parameters~\cite{damour:1991rd,hulse:1974eb,taylor:1982zz,Taylor:1989sw}. In this section, we discuss the use of precise orbital parameter data to place constraints on the Lorentz-violating Einstein-\ae ther theory. In GR energy is carried away at quadrupolar order due to propagation of tensor modes whereas in this theory (and many of other modified theories of gravity), one usually finds radiation from extra scalar and vector modes which are responsible for energy loss at dipole order, i.e., $-1$PN order (c.f. the term proportional to $\mathcal{E}$ in Eq.~\eqref{Pdot-AE}) as compared to GR. Hence, energy is radiated faster than what is predicted in GR. This results in a decrease in the orbital separation and orbital period ($P_{b}$) of the binary.  The modified orbital period decay rate ($\dot{P_{b}}$) relates to the total energy of the binary, i.e., Eq.~\eqref{Pdot-AE} via
\begin{align}\label{eq:PdotEdot}
\frac{\dot{P_{b}}}{P_{b}} = -\frac{3}{2}\frac{\dot{E_{b}}}{E_{b}}\,,
\end{align} 
suggesting a strong dependence of $\dot{P_{b}}$ on the sensitivity of the NS~\cite{blas:2010hb}. Since the GR predictions agree with the observed value of $\dot{P_{b}}$ within observational uncertainty, this allows for stringent constraints to be placed on \ae ther theory.

\subsection{Observations}

Following from Eqs.~\eqref{Pdot-AE} (with LV terms set to zero)
and \eqref{eq:PdotEdot},
one can relate the post-Keplerian parameter $\dot{P_{b}}$ in GR to the Keplerian parameter $P_{b}$ via \cite{Yagi:2013ava}
\begin{align}\label{PbdotGR}
\left(\frac{\dot{P_{b}}}{P_{b}}\right)_{\text{GR}} = -\frac{384 \pi}{5} 2^{2/3}\left(\frac{\pi (m_1+m_2)}{P_{b}}\right)^{5/3}\frac{m_1 m_2}{(m_1+m_2)^{2}} \frac{1}{P_{b}}.
\end{align}
Here $m_1$ and $m_2$ are the masses in the binary system.
In principle, if we can measure the masses, orbital periods and the orbital period decay rates with some uncertainty and find that they are consistent with GR predictions, then we can place constraints on Einstein-\ae ther theory. In this section we focus on data from the measurements of Keplerian and post Keplerian parameters of four different pulsar systems PSR J1738+0333, PSR J0348+0432, PSR J1012+5307 and PSR J0737-3039 (Table~\ref{tab:pulsarsystem}) and a stellar triple system~\cite{voisin_2020}. The first three are pulsar-white dwarf binaries in orbits with $\mathcal{O}(10^{-7})$ eccentricity and 8.5-hour period, 0.17 eccentricity and 4.74-hour period and $\mathcal{O}(10^{-7})$ eccentricity and 14.5-hour period respectively. The fourth is the double pulsar binary system with 0.088 eccentricity and 2.45-hour period. 
Because of the small eccentricity of these systems, 
we will ignore it in the following, i.e. we will consider quasi-circular binaries.

\setlength{\tabcolsep}{8pt}
\renewcommand{\arraystretch}{1.3}
\begin{table}[ht]
	\caption{
    Orbital parameters as measured for the binary systems studied in this paper. Table shows the estimated values of the parameters and the 1-$\s$ uncertainty in the last digits in parentheses. Here, $\dot{P_{b}}^{\text{obs}}$ is the observed value of $\dot{P_{b}}$.}
	\fontsize{10}{12}\selectfont 
		\begin{tabular}{ | c  | c  | c  | c  | c  |} 
			\hline 
			Pulsar System&  $m_{1} (M_{\odot})$ & $m_{2} (M_{\odot})$ & $P_{b}$ (days) &  $\dot{P_{b}}^{\text{obs}}$\\
			\hline \hline
			PSR J1738+0333\cite{Freire_2012} & $1.46^{+0.06}_{-0.05}$ & $0.181^{+0.008}_{-0.007}$ & $0.3547907398724(13)$ & $-25.9(3.2)\times 10^{-15}$\\ 
			\hline
			PSR J0348+0432 \cite{Antoniadis_2013}& 2.01(4) & 0.172(3) & $0.102424062722(7)$ & $-0.273(45)\times 10^{-12}$ \\ 
			\hline
			PSR J1012+5307 \cite{10.1046/j.1365-8711.1998.01634.x} \cite{10.1111/j.1365-2966.2009.15481.x}& $1.64(0.22)$ & $0.16(0.02)$ & $0.60467272355(3)$ & $-1.5(1.5) \times 10^{-14}$\\ 
			\hline
			PSR J0737-3039 \cite{Kramer:2006nb} & 1.3381(7) & 1.2489(7) & 0.10225256248(5) & $-1.252(17)\times 10^{-12}$\\
			\hline
		\end{tabular}
		\label{tab:pulsarsystem}
\end{table}
\subsection{Parameter Estimation and Bayesian Analysis}

Our goal is to constrain the theory parameters using measurements of $\dot{P_{b}}$. We discuss briefly the Bayesian formalism with Markov-Chain Monte-Carlo (MCMC) exploration used to calculate the posteriors on the model parameters [c.f. Sec. \ref{sec:bayes}]  and derive robust constraints. 

For the parameter estimation, we need the expression for the orbital period decay, which depends on both the relative velocity of the binary constituents $v_{21}$ and the center-of-mass velocity $w$ of the binary's center of mass with respect to the \ae ther field. A natural choice for the \ae ther field direction is provided by the cosmic microwave background
(i.e. the \ae ther is expected to be approximately aligned with
the cosmological background time direction). In this case, a typical value for the center-of-mass velocity is $w \sim 10^{-3}$~\cite{foster:2007gr}, which for binary pulsar observations is of the same order as $v_{21}$. If so, the 
$w$-dependent corrections in the rate of change of the binding energy [Eq.~(\ref{Pdot-AE})]
and orbital period [Eq.~(\ref{PbdotAE})] enter at the same PN order (0PN) as the quadrupole emission terms of GR, but multiplied by either $(s_1-s_2)$ 
or powers of it. As such, these
$w$-dependent corrections   are negligible for both
white dwarf-pulsar systems (for which the dominant term is the $-1$PN dipole emission) and also for the relativistic double pulsar system  (for which
$s_1-s_2\approx 0$ as a result of the similar pulsar masses, which kills both the dipole emission and the  
$w$-dependent corrections to quadrupole emission). Therefore, our results are independent of the exact value of $w$ as long as that is of order
$w \sim 10^{-3}$ or smaller~\cite{Yagi:2013ava}

From the above assumption and using Eqs.~\eqref{eq:PdotEdot} and~\eqref{Pdot-AE}, the orbital period decay rate in Einstein-\ae ther theory is a function of the individual masses $(m_{1},m_{2})$~\footnote{These are the active masses, whose fractional difference
from the ``real'' masses  $(\tilde{m}_{1},\tilde{m}_{2})$
is of the order of the sensitivities and thus negligible. In the following we will therefore typically identify  $(m_{1},m_{2})$ and $(\tilde{m}_{1},\tilde{m}_{2})$. Note that this could however introduce correlations not captured by our sufficient statistics approach, but as we show, even large correlations would have little impact on the results.}, pulsar radii $(R_{\star,1},R_{\star,2})$, orbital period ($P_{b}$) and coupling constants $(\a_{1},\a_{2},c_{\omega})$ as shown below 
\begin{align}\label{PbdotAE}
\nonumber\frac{\dot{P_{b}}}{P_{b}} = &\frac{1}{5(m_1 + m_2)^4 P_{b}\sqrt{\frac{\a_1}{(\a_1 - 8\a_2)}}\sqrt{\frac{-c_{\omega}}{\a_1}} \a_1^3 c_{\omega}^2} \Biggl\{3\,2^{8/3} \pi \Biggl(\frac{\pi (m_1 + m_2)}{P_b}\Biggr)^{5/3}\\
\nonumber&m_1 m_2 \biggl[-\frac{1}{12}\Biggl(5 \a_1 c_{\omega} (m_1 + m_2)^2 (s_1 - s_2)^2 \\
\nonumber&\Biggl(2^{5/6} \a_1^2 (\a_1 + 8) \sqrt{\frac{\a_1}{(\a_1 - 8\a_2)}} - 2^{13/3} \sqrt{\frac{-c_{\omega}}{\a_1}} c_{\omega} (\a_1 - 8\a_2)\Biggr)\\
\nonumber&\Biggl(\frac{P_b}{\pi m}\Biggr)^{2/3}\Biggr) +\Biggl((s_1-1)(s_2-1)\Biggr)^{2/3}\Biggl((\a_1 + 8)\a_1^3 \\
\nonumber&\Biggl(-4 c_{\omega}^2 (m_1 + m_2)^2 \sqrt{\frac{-c_{\omega}}{\a_1}} + \sqrt{2} \a_1 (m_1 s_2 + m_2 s_1)^2\Biggr) \sqrt{\frac{\a_1}{(\a_1 - 8 \a_2)}} +\\
&\frac{2 \sqrt{\frac{-c_{\omega}}{\a_1}} (\a_1 - 8 \a_2)^2 c_{\omega}^2 ((m_1 + m_2) \a_1 + 8 m_1 s_2 + 8 m_2 s_1)^2}{3}\Biggr) \Biggr] \Biggr\}\,,
\end{align}
where $s_{1}$ and $s_{2}$ are functions of $\mathcal{C}$ and coupling constants. One may worry that the terms inside the square roots in the above expression may be negative, leading to a complex orbital decay rate, but this is not the case because when $\alpha_1 > 0$, then  $c_{\omega}\leq -\alpha_1/2$, while when $\alpha_1 < 0$, then $c_{\omega}\geq -\alpha_1/2$.
As noted in Fig.~\ref{fig:universal_eos}, sensitivities are independent of the EoS. Here we choose to work with the Tolman VII EoS since it gives stable analytic solutions for the sensitivities.

There are some phenomenological constraints on the \ae ther coupling constants as discussed in Sec.~\ref{sec:AE1}, i.e., $|\alpha_1| \lesssim 10^{-4}$, $|\alpha_2| \lesssim 10^{-7}$ (Solar system constraints), $\alpha_1 < 0$, $\alpha_1 < 8 \alpha_2 <0$ and $c_\omega > -\alpha_1/2$ (positive energy, absence of vacuum Cherenkov radiation and gradient instabilities) and $c_{\sigma} \lesssim 10^{-15}$ (GW constraint).\footnote{Note that
we cannot use existing bounds on 
$\hat{\alpha}_1$~\cite{shao:2012eg,Will:2014kxa}  and $\hat{\alpha}_2$~\cite{shao:2013wga, shao:2012eg, Will:2014kxa} as priors. This is because those quantities depend on the derivatives of the sensitivities (c.f.~\ref{app:EIH}), which are currently unknown.
} Using these pre-existing constraints and by determining if the estimated value of $\dot{P_{b}}$ lies within the range $\dot{P_{b}}^{\text{obs}} \pm \delta\dot{P_{b}}^{\text{obs}}$ (Table~\ref{tab:pulsarsystem}), we determine the consistency of points in the parameter space with observations.

One important point is that we are not using the pulsar timing data directly~\cite{Anderson_2019}, but instead we are using existing constraints on $\dot{P_{b}},~m_1$ and $m_2$ derived from the primary pulsar timing data as a {\em sufficient statistic}. Unfortunately, the published results only quote values for the individual parameters and their uncertainties, so we do not have access to the joint posterior distributions. For simplicity we assume that the parameter correlations are negligible. To check the impact of this assumption, we compared results with zero correlations with a case with 90\% correlation between the parameters, and found that it only changed the results by a maximum of 17\%. 

\subsubsection{Bayesian Analysis}\label{sec:bayes}

We are interested in constructing a posterior distribution on a set of model parameters $\vec{\lambda}$ = ($m_1$, $m_2$, $P_{b}$, $R_{\star,1}$, $R_{\star,2}$, $\a_{1}$, $\a_{2}$, $c_{\omega}$) and using an MCMC algorithm to explore the parameter space. According to Bayes' theorem, the probability density for parameters $\vec{\lambda}$ given data D and hypothesis H (the theory) is
\be
P(\vec{\lambda}|\text{D},\text{H}) = \frac{P(\text{D}|\vec{\lambda} , \text{H}) P(\vec{\lambda}|\text{H})}{P(\text{D}|\text{H})} \,,
\ee 
where $P(\vec{\lambda}|\text{H})$ is called the prior which represents the state of knowledge about the parameters before we analyze the data. $P(\text{D}|\vec{\lambda},\text{H})$ is called the likelihood which describes the probability of measuring data D given the model H and a set of parameters $\vec{\lambda}$. $P(\text{D}|\text{H})$ is called the model evidence which represents the overall normalization factor. In practice it is better to work with log probability densities to better cover the dynamic range of the densities. 

We assumed uniform priors on $\a_{2}$ and $c_{\omega}$ such that  $-4\times 10^{-7} \leq \a_{2} \leq 4\times 10^{-7} $ and $-10^{5} \leq c_{\omega} \leq 10^{5} $ and a Gaussian prior for $\a_{1}$, $m_1$, $m_2$, $\dot P_b$ with mean and standard deviation given by the existing bounds listed in Table~\ref{tab:pulsarsystem}. We use Gaussian priors on $R_{\star,1}$ and $R_{\star,2}$ with mean and standard deviation given $12.4 \pm 1.1$km based on LIGO and NICER measurements~\cite{Coughlin_2019}. While these bounds are derived assuming GR, the corrections due to LV effects are sub-dominant compared to those impacting $\dot P_b$ (c.f. e.g. footnote~$\mathcal{P}$). Using lunar laser ranging experiments, the bounds on $\alpha_1$ were obtained to be $\a_{1} =  (-0.7 \pm 0.9) \times 10^{-4}$~\cite{muller:2005sr} (c.f. also Sec.~\ref{sec:AE1}).

The log likelihood function is
\begin{equation}
\ln (P(\text{D}|\vec{\lambda},\text{H})) \propto  -\frac{1}{2}\frac{\left(\left(\dot{P_{b}}/P_{b}\right)^{\text{obs}} - \left(\dot{P_{b}}/P_{b}\right)^{\text{th}}\right)^{2}}{\sigma^2_{(\dot{P_{b}}/P_{b})}} \,,
\end{equation}
where $(\dot{P_{b}}/P_{b})^{\text{th}}$ is the theoretically predicted value of $(\dot{P_{b}}/P_{b})$ from the model. With the likelihood and the priors in place, we can find the posterior using a MCMC algorithm.

We start the MCMC simulation near the mean values for the model parameters, calculate the posterior and iterate through these steps. Model parameters are allowed to explore the entire range of parameter space and that gives the joint posterior distribution on all parameters $\vec{\lambda}$. For the proposal distribution we use the prior distribution for a certain set of parameters, and a relative jump from the current position for the remaining. Proposed jumps are accepted or rejected based on the Metropolis-Hastings acceptance probability
\be\label{mh_ratio}
\text{H} = \text{min}\left( \frac{P(\vec{\lambda}_\mathrm{new}|H) P(D|\vec{\lambda}_\mathrm{new},H) Q(\vec{\lambda}_\mathrm{old}|\vec{\lambda}_\mathrm{new})}{P(\vec{\lambda}_\mathrm{old}|H) P(D|\vec{\lambda}_\mathrm{old},H) Q(\vec{\lambda}_\mathrm{new}|\vec{\lambda}_\mathrm{old})}, 1 \right) \,.
\ee
A random number $u \sim U[0,1]$ is drawn, and if $\text{H} > u$ the proposed jump is accepted, otherwise it is rejected. This process is repeated multiple times to ensure convergence.\\
\begin{figure}
    \includegraphics[scale=0.59]{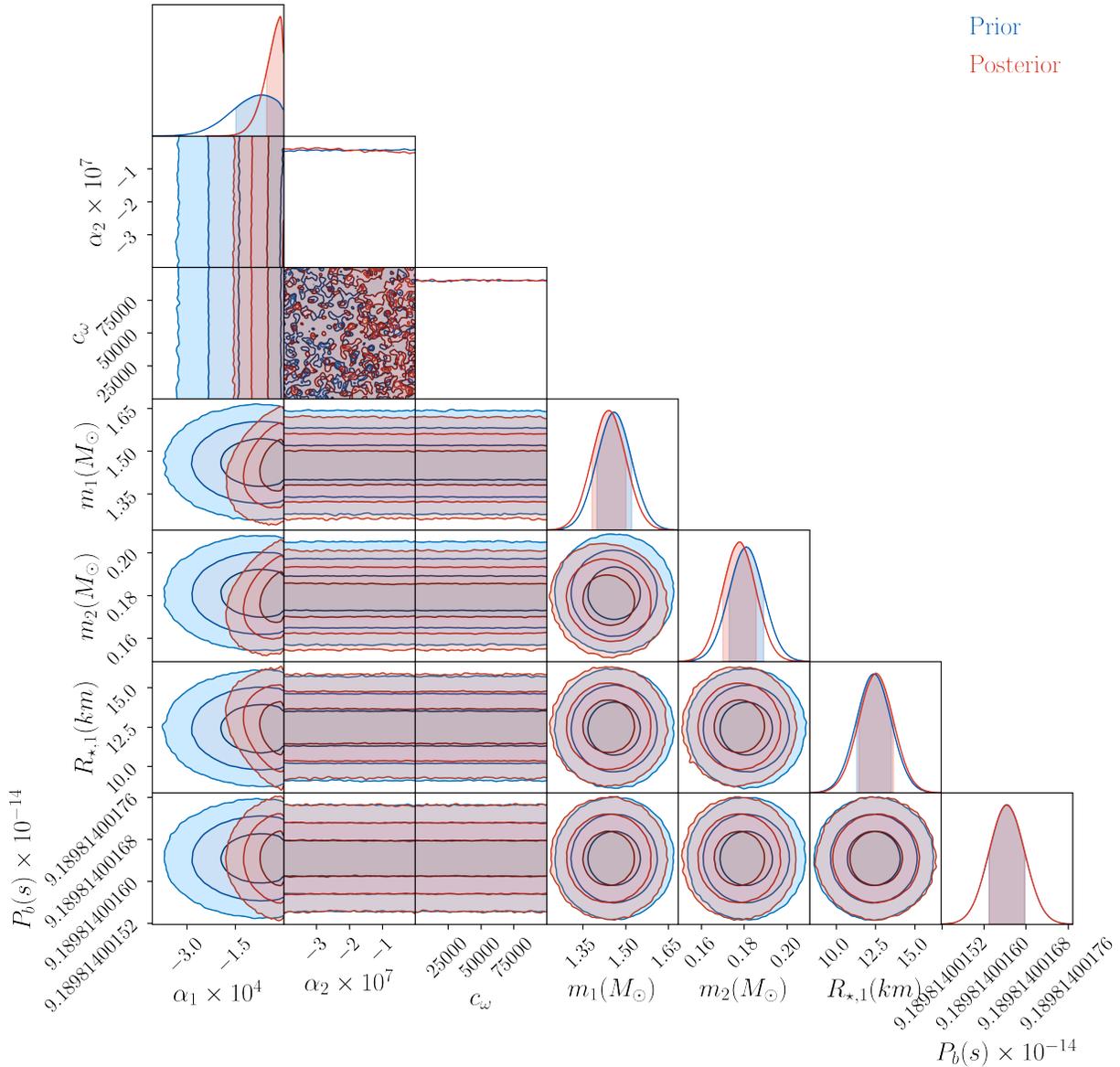}
    \caption{Prior and posterior distribution on the model parameters $\vec{\lambda}$ from $\dot{P_{b}}$ constraints for PSR J1738+0333.
    The pre-existing constraints from solar system, Big-Bang nucleosynthesis and stability requirements are applied to uniform priors shown in blue where, for $\a_{1}$ negative it results in a negative $\a_{2}$ and positive  $c_{\omega}$. The posterior distributions depend on the $\dot{P_{b}}$ constraints for PSR J1738+0333. The three shades of contours in the prior and posterior distribution in the off-diagonal cross-correlation panels represent 1-$\s$, 2-$\s$ and 3-$\s$ uncertainty on model parameters starting from the center (we only show 1-$\s$ shaded regions for the one-dimensional marginal distributions). There is a small dip at very small magnitudes of $\a_1$, as that is the region where the pre-existing constraints come into play, while for larger magnitudes the constraints on $\a_1$ are automatically satisfied. Observe that the value of $\a_{1}$ is further constrained by a factor of 2 compared to existing solar system constraints (prior).} 
    \label{fig:pulsar1}
\end{figure}

\begin{figure}
    \centering
    \includegraphics[scale=0.9]{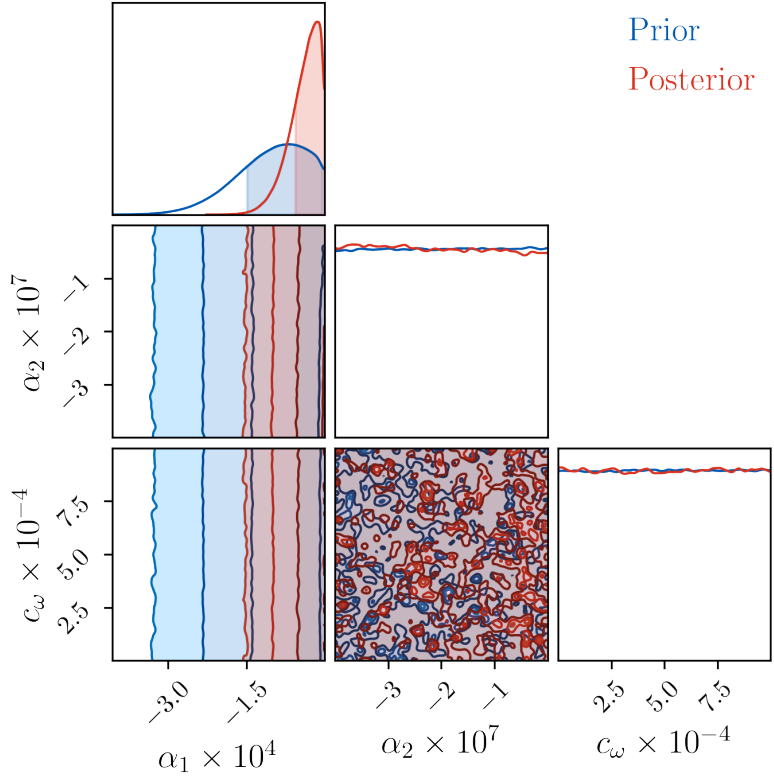}
    \caption{Joint prior and posterior distribution on $\a_1$, $\a_2$ and $c_\omega$ from $\dot{P_{b}}$ constraints for all four pulsars listed in (Table \ref{tab:pulsarsystem}) The constraint on $\a_{1}$ is improved by a factor of 2.}
    \label{fig:joint_pulsars}
\end{figure}
We begin by considering a single observation from the pulsar-white dwarf system PSR J1738+0333. Since the sensitivity of a white dwarf (WD) is negligible compared to the NS we can set $s_2$ = $s_{\mbox{\tiny{WD}}} = 0$ (thus $R_{\star,2}$ is excluded from $\vec \lambda$) but for a double pulsar binary we should have $s_{2} \neq 0$. Figure \ref{fig:pulsar1} shows the prior and posterior distribution on the model parameters. We are recovering our priors on the masses and radii, given that these are well constrained as can be noted from Table \ref{tab:pulsarsystem}. The distribution on $\alpha_1$ and $\alpha_2$ is such that $\alpha_1 < 0 $ and $\alpha_2 < 0$ from existing constraints. This pulsar system further constrains the value of parameter $\a_{1}$ by approximately a factor of 2 while the coupling constants $\a_{2}$ and $c_{\omega}$ remain unconstrained. Notice that the posteriors on $\alpha_2$ and $c_\omega$ are flat and very similar to the priors. Therefore, one cannot model them as Gaussian and construct confidence region, as no information is gained for the values of these parameters. 

We then consider constraints on the coupling parameters by stacking all four different binary systems from Table~\ref{tab:pulsarsystem} and computing the joint constraints (Fig.~\ref{fig:joint_pulsars}). These joint constraints also restrict the region of $\a_{1}$ by a factor of 2 better than the existing constraints.

\begin{figure}
    \centering
    \includegraphics[scale=0.46]{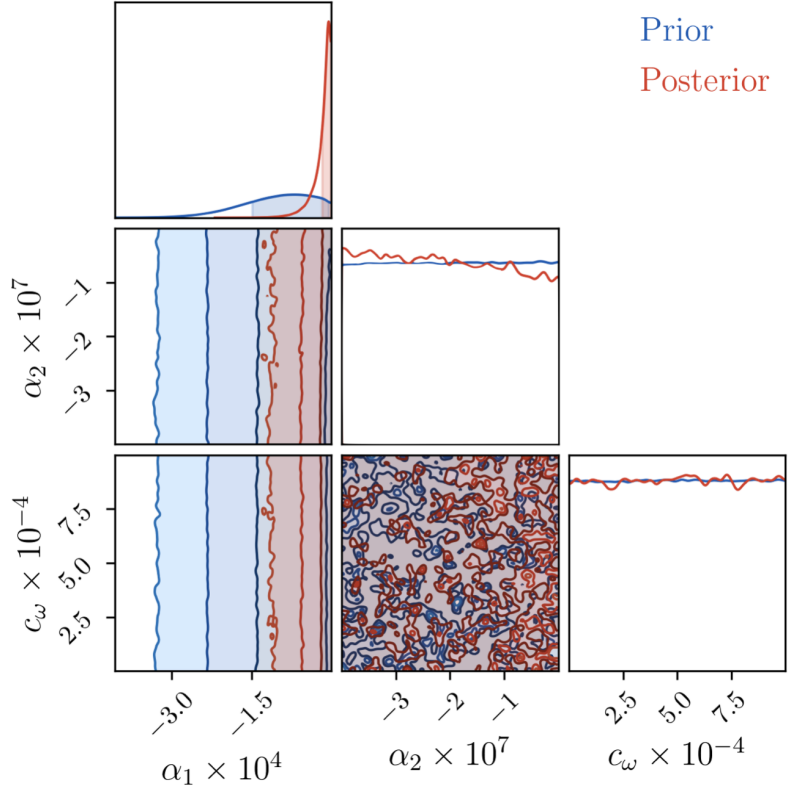}
    \caption{Prior and posterior distribution on the model parameters $\a_1$, $\a_2$ and $c_\omega$ from $\dot{P_{b}}$ constraints for PSR J0348+0432, in a scenario where the observational uncertainties tighten by a factor of 10 and the value of $\dot{P_{b}}$ matches the GR prediction. It shows that the GR values are favoured and the value of $\a_{1}$ is very closely centered around zero.}
    \label{fig:pulsar1_redby10GR}
\end{figure}

\begin{figure}
    \centering
    \includegraphics[scale=0.9]{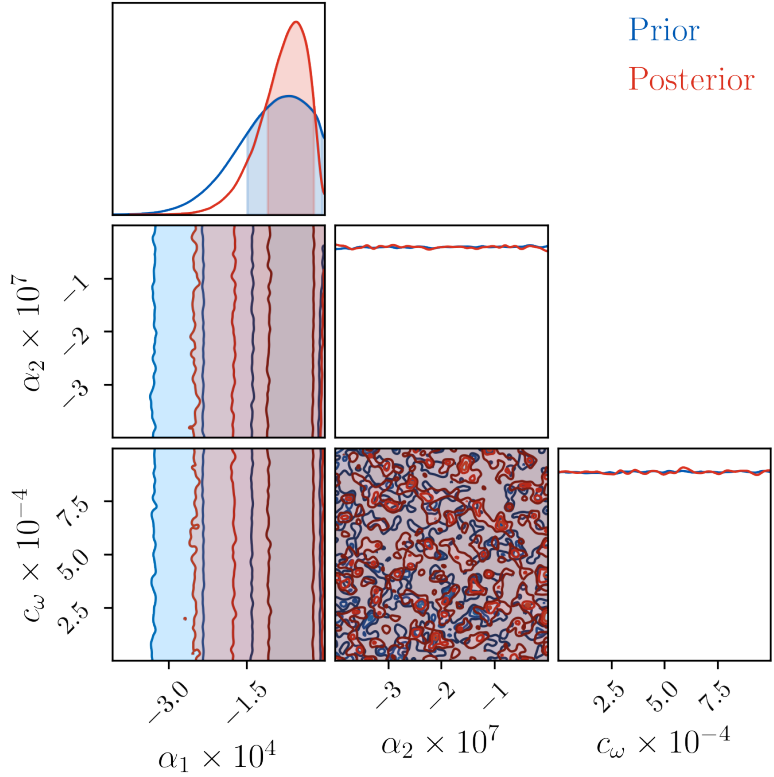}
    \caption{Prior and posterior distribution on the model parameters $\a_1$, $\a_2$ and $c_\omega$ from $\dot{P_{b}}$ constraints for PSR J0348+0432, in a scenario where the uncertainties in measurements are reduced by a factor of 10, and the value of $\dot{P_{b}}$ stays at the currently observed value. The 1-$\sigma $ uncertainty shows that $\a_1=0$, i.e., the GR value is disfavoured. It can also be noted from Table~\ref{tab:alpha1_bounds} that posterior does not include $\a_1 = 0. $ }
    \label{fig:pulsar1_redby10}
\end{figure}

In the coming years we expect to have more observations, 
as the sensitivities of radio telescopes will improve as a result of
larger collecting areas (e.g. the  Square Kilometre Array (SKA) project~\cite{Bull:2018lat,Braun:2019gdo}), which will
allow for discovering more pulsars. Moreover
the longer observation time ($T$) will reduce the error in measurements of $\dot{P_b}$ by $T^{-5/2}$~\cite{1976ApJ...205..580B}, allowing for more precise measurements of the orbital parameters. 
One may wonder, whether we can get tighter constraints from finding $N$ similar systems or a single system measured with higher SNR (signal-to-noise ratio). The SNR$^2$ grows linearly with number of sources $N$ and the observing time $T$, and quadratically with the effective collecting area of the radio telescope. Significant improvements in the measurement sensitivity are more likely to come from some combination of larger telescopes and additional observing time than from discovering large numbers of systems similar to those known.
Figure~\ref{fig:pulsar1_redby10GR} illustrates the kind of bounds we will get for a PSR J0348+0432 system if ${\dot{P_{b}}}^{\text{obs}}$ matches the GR prediction ${\dot{P_{b}}}^{\text{GR}}$ and the uncertainties are tightened by a factor of 10. The improved constraints on ${\dot{P_{b}}}$ translate directly into similarly improved bounds on $\a_{1}$. We also considered an alternative scenario, in which the uncertainties in ${\dot{P_{b}}}$ improved by a factor of 10, but stayed centered on the current observed value. As shown in Figure~\ref{fig:pulsar1_redby10}, this leads to a value for $\a_{1}$ bounded away from zero. 
In other words, in a scenario where the observed period derivative stays at the current value while the uncertainty drops by a factor of ten, we would find that Einstein-\ae ther theory would be favored over GR!

\subsection{Constraints from the triple system}

\begin{figure}[h!]
    \centering
    \includegraphics[scale=0.9]{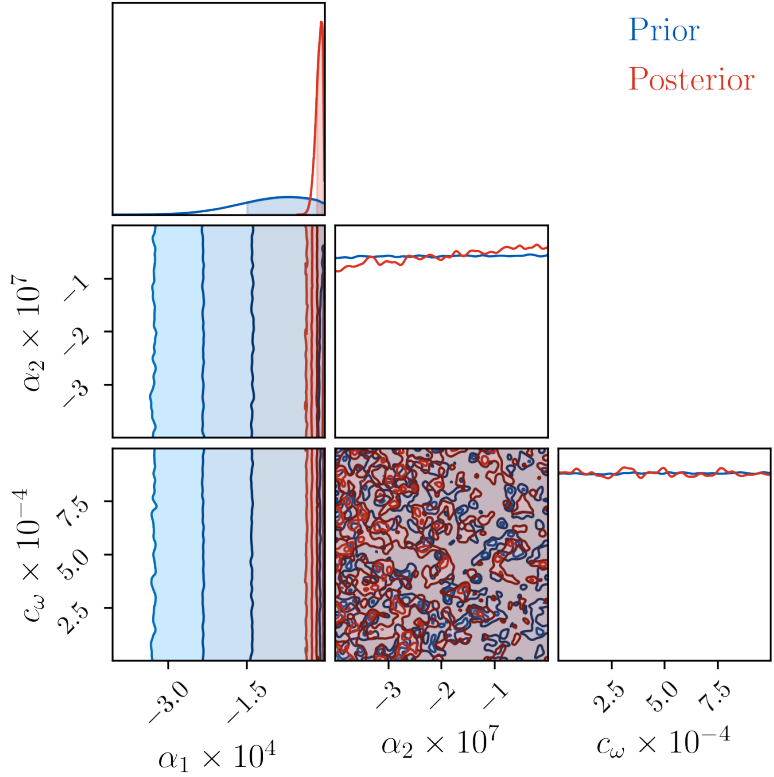}
    \caption{Joint prior and posterior distribution on the model parameters $\vec{\lambda}$ from $\dot{P_{b}}$ constraints for all four pulsars listed in (Table~\ref{tab:pulsarsystem}) and stellar triple system. Observe that inclusion of stellar triple system improves the constraints and parameter $\a_{1}$ is now constrained by a factor of 10 better than lunar laser ranging experiments.}
    \label{fig:joint_pulsar_triple}
\end{figure}

Next we have constraints coming from a pulsar in a stellar triple system PSR J0337+1715 consisting of an inner millisecond pulsar-white dwarf binary and a second white dwarf (WD) in an outer orbit~\cite{archibald:2018oxs}. Due to the gravitational pull of the outer WD, the pulsar and the inner WD experience accelerations that differ fractionally. If the strong equivalence principle is violated (as a result of the sensitivities), the triple system constrains the fractional acceleration difference parameter $\delta_a$ to $(+0.5\pm1.8)\times 10^{-6}$~\cite{voisin_2020}. The relation between $\delta_a$ and the sensitivity parameter $\sigma_{\rm pulsar}$ (before rescaling) in Einstein-\ae ther theory is~\cite{will:2018ont,Barausse:2019yuk}
\begin{equation}\label{delta_eq}
|\delta_a|=\left\vert\frac{\sigma_{\rm pulsar}}{1+\sigma_{\rm pulsar}/2}\right\vert\approx |\sigma_{\rm pulsar}|\,,
\end{equation}
as can be obtained directly from Eq.~\eqref{eq:acceleration} (in the Newtonian limit).

We use MCMC simulations in Bayesian analysis similar to that for the $\dot{P_{b}}$ constraint and with the likelihood
\begin{equation}
P(\text{D}|\vec{\lambda},\text{H}) \propto  \exp\left({-\frac{1}{2}\frac{\left(\sigma_{\rm pulsar}^{\text{obs}} - \sigma_{\rm pulsar}^{\text{ th}}\right)^{2}}{\sigma^2_{(\sigma_{\rm pulsar})}}}\right)\,,
\end{equation}

where $\sigma_{\rm pulsar}^{\text{obs}} = (+0.5\pm1.8)\times 10^{-6}$ from Eq.~\eqref{delta_eq} and $\sigma_{\rm pulsar}^{\text{th}}$ is given by Eq.~\eqref{sens_method1}, to constrain the model parameters. Figure \ref{fig:joint_pulsar_triple} shows the joint pulsar and triple system constraints on the model parameters $\vec{\lambda}$ assuming uniform distribution in $\a_{2}$ and $c_{\omega}$. The 95\% upper limit on $\alpha_1$, which was $-2.4 \times 10^{-4}$ (from the prior constraints) has now shifted to $\alpha_1$ = $-2.4 \times 10^{-5}$. It shows that the preferred frame parameter $\a_{1}$ is constrained by a factor of 10 better than the lunar laser ranging experiments.

\begin{table}
\caption{Our bounds on $\a_{1}$ from different pulsar systems shown in Figs.~\ref{fig:pulsar1}--\ref{fig:joint_pulsar_triple} with 1-$\s$ uncertainity. The first half shows the bounds from existing measurements, while the second half shows projected future bounds assuming that the measurement error on $\dot P_b^\mathrm{obs}$ reduces by a factor of 10 with the central value of $\dot P_b^\mathrm{obs}$ at the GR predicted value ($- 27.3\times10^{−14}$) and at the current measured value ($−25.3\times10^{−14}$).}
\label{tab:alpha1_bounds}
\begin{center}
\begin{tabular}{|c|c|}
    \hline
    Pulsar System & $\a_{1}$ \\
    \hline
    PSR J1738+0333 & $(-3.975 \pm 2.968)\times 10^{-5}$ \\
    Joint binary system & $(-4.073 \pm$ 2.936)$\times 10^{-5}$ \\ 
    Joint binary + triple system & ($-1.111$ $\pm$ 0.674)$\times 10^{-5}$ \\
    \hline
    PSR J0348+0432 ($\dot{P_{b}}^{\text{obs}} = - 27.3\times10^{−14}$) & ($-8.119$ $\pm$ 4.622)$\times 10^{-5}$ \\
    PSR J0348+0432 ($\dot{P_{b}}^{\text{obs}} = −25.3\times10^{−14}$) & $(-1.729 \pm$ 1.805)$\times 10^{-5}$ \\
    \hline
\end{tabular}
\end{center}
\end{table}
Table~\ref{tab:alpha1_bounds} shows bounds on $\a_1$ from binary and triple systems mentioned in this paper. The data from joint binary + triple system allows us to put a stringent constraint on $\a_{1}$, which is an order of magnitude stronger than the bounds from lunar laser ranging experiments~\cite{will_2018,muller:2005sr}.

\section{Conclusions}\label{sec:concl}

We have investigated Einstein-\ae ther theory in the context of binary pulsars and NSs. We have recalculated the sensitivities in the regime of coupling parameter space that still survives after the recent measurement of the speed of GWs. This required the development of a new post-Minkowskian approach that allows for stable numerical evaluation of the sensitivities, in addition to the derivation a closed form analytic solution for the Tolman VII EoS. We used these results to place a constraint on certain coupling constants of Einstein-\ae ther theory using Bayesian analysis of binary pulsar observations, including recent observations on the triple system. We find that these data allows for constraints on a certain combination of the coupling constants, $\alpha_1$, of ${\cal{O}}(10^{-5})$, improving current Solar System constraints by one order of magnitude. 

The work carried out here opens the door to several avenues for future research. One such avenue is to use gravitational wave data directly to place constraints on Einstein-\ae ther theory, now that the sensitivites have been analytically calculated. This can be done today to leading post-Newtonian order in the inspiral, and it remains to be seen whether it is enough to lead to interesting constraints. To include the very late inspiral and merger phase, numerical simulations of coalescing NSs would have to be carried out in Einstein-\ae ther theory. However, since the parameter space of the theory is already quite well constrained, it is not clear whether stronger bounds can be achieved with gravitational wave data. 

Another avenue for future research concerns computing sensitivities for black holes. This has been done in khronometric theory~\cite{ramos} but not yet in Einstein-\ae ther theory. Once the black hole sensitivities are in hand, and assuming they do not vanish, one could use the existing GW data for binary black hole mergers to constrain Einstein-\ae ther theory, including the dipole radiation effect in the gravitational waveform.

One more avenue for future work would be along the lines of improving the analysis in this paper by directly analyzing binary pulsar data and carrying out a parameter estimation and model selection study with a GR and a non-GR timing model. For this, it would be ideal to compute the derivative of the NS sensitivities that enter in the conservative post-Keplerian parameters, such as the periastron precession and Shapiro time delay. 

\section*{Acknowledgements} 
We would like to thank Clifford Will and Ted Jacobson for many insightful discussions, which motivated part of this work. TG and NJC acknowlege the support of NASA EPSCoR grant MT-80NSSC17M0041 and NSF grant PHY-1912053. KY acknowledges support from 
NSF Grant PHY-1806776, NASA Grant No.~80NSSC20K0523, a Sloan Foundation Research Fellowship and the Owens Family Foundation. 
KY would like to also thank the support by the COST Action GWverse CA16104 and JSPS KAKENHI Grants No. JP17H06358.
NY acknowledges support from NASA grant No.~NNX16AB98G, No.~80NSSC17M0041, No.~80NSSC18K1352 and NSF grant PHY-1759615. 
EB and MHV acknowledge financial support provided under the European Union's H2020 ERC Consolidator Grant
``GRavity from Astrophysical to Microscopic Scales'' grant agreement no. GRAMS-815673.


\appendix

\section{Modified EIH technique}\label{app:EIH}
In this Appendix, we start from the 1PN acceleration \eqref{eq:acceleration}
and analyze the effect of the 1PN conservative dynamics  on the orbital parameters of a binary of compact objects.
We will follow the osculating-orbits technique of 
\cite{will:2018ont}, which will lead us to amend the calculation of the
preferred frame parameters $\hat{\alpha}_1$ and $\hat{\alpha}_2$ presented in 
\cite{Yagi:2013ava}. In doing so, we will also correct a few typos that we found in the expressions of \cite{will:2018ont}.\footnote{We have double checked the correctness of our expressions with the authors of \cite{will:2018ont}.}

The relative acceleration between the two gravitating bodies is obtained by simply letting
\begin{align}
\bm{a}=\frac{d\bm{v}_1}{dt}-\frac{d\bm{v}_2}{dt},
\end{align}
while the position of the center of mass is not accelerated. Thus, we can set $\dot{\bm{X}}=\bm{X}=0$ at Newtonian order without any loss of generality, getting 
\begin{align}
&\bm{x}_1=\left(\frac{m_2}{m}+{\cal O}(\epsilon)\right)\bm{x},\\
&\bm{x}_2=-\left(\frac{m_1}{m}+{\cal O}(\epsilon)\right)\bm{x},
\end{align}
where $\epsilon\sim m/r \sim v_{21}^2$ is a book-keeping parameter that counts PN order.

Here the acceleration of every individual body is given by \eqref{eq:acceleration}. Hereinafter we will borrow the notation of \cite{will:2018ont} and thus we define
\begin{align}
m=m_1+m_2, \quad \eta=\frac{m_1m_2}{m^2},\quad \Delta=\frac{m_2-m_1}{m},
\end{align}
and the functions of the sensitivities 
\begin{align}
\nonumber &\mathcal{G}=\mathcal{G}_{12}, \quad \mathcal{B}_+=\mathcal{B}_{(12)}, \quad \mathcal{B}_-=\mathcal{B}_{[12]},\quad  \mathcal{D}=\frac{m_2}{m}\mathcal{D}_{122}+\frac{m_1}{m}\mathcal{D}_{211},\\
& \mathcal{E}=\mathcal{E}_{12},\quad \mathcal{A}^{(n)}=\left(\frac{m_2}{m}\right)^n \mathcal{A}_1-\left(-\frac{m_1}{m}\right)^n \mathcal{A}_2.
\end{align}

Using this, the relative acceleration can be written in a compact form
\begin{align}\label{eq:force}
\bm{a}=\bm{a}_{\rm L}+\bm{a}_{\rm PF},
\end{align}
where we have separated the purely local contributions and those coming from preferred frame effects. The former reads
\begin{align}
\bm{a}_{\rm L}=\frac{m}{r^2}\left[\bm{n}\left(\hat{A}_1 v_{21}^2 +\hat{A}_2 \dot{r}^2+\hat{A}_3 \frac{m}{r}\right)+\dot{r}\hat{B}\bm{v}_{21}\right],
\end{align}
where
\begin{align}
&\bm{v}_{21}=\dot{\bm{x}}_2-\dot{\bm{x}}_1,\\
&\hat{A}_1=\frac{1}{2}\left[\mathcal{G}(1-6\eta)-3\mathcal{B}_+-3\Delta\mathcal{B}_--\eta(\mathcal{C}_{12}+2\mathcal{E})+\mathcal{G}\mathcal{A}^{(3)} \right],\\
&\hat{A}_2=\frac{3\eta}{2}(\mathcal{G}+\mathcal{E}),\\
&\hat{A}_3=\mathcal{D}+\mathcal{G}\left[2\eta \mathcal{G}+3 \mathcal{B}_++\eta (\mathcal{C}_{12}+\mathcal{E})+3\Delta \mathcal{B}_-\right],\\
&\hat{B}=\mathcal{G}(1-2\eta)+3\mathcal{B}_++3\Delta \mathcal{B}_-+\eta\mathcal{G}+\mathcal{G}\mathcal{A}^{(3)}.
\end{align}

These expressions agree with those of \cite{will:2018ont}. However, we find a difference in the acceleration due to preferred frame effects 
\begin{align}
\bm{a}_{\rm PF}&=\frac{m}{r^2}\left\{ -\bm{n}\left[\left(\frac{\hat\alpha_1}{2}+2{\cal G}{\cal A}^{(2)}\right) (\bm{\omega}\cdot \bm v_{21})+\frac{3}{2}\left(\hat\alpha_2 + {\cal G}{\cal A}^{(1)}\right)(\bm{\omega}\cdot\bm{n})^2\right]\right. \\
\nonumber&\left .-\bm{\omega}\left[\frac{\hat\alpha_1}{2}(\bm n \cdot \bm v_{21})+\hat{\alpha}_2(\bm{n}\cdot \bm \omega)\right]+ {\cal G}{\cal A}^{(2)} \bm{v}_{21}(\bm n \cdot \bm \omega)\right\}-\frac{m\omega^2}{2r^2} \left({\cal C}_{12}+{\cal G}{\cal A}^{(1)}\right)\bm{n}.
\end{align}
where we have already specified the generic boost velocity $\boldsymbol{w}$ in \eqref{eq:acceleration} to match the velocity of the preferred frame $\bm \omega$.

This differs from the result of \cite{will:2018ont} in a sign multiplying the first whole line, as well as in the last term, which is absent in \cite{will:2018ont}. Here we have defined the following compact-body effective PPN parameters
\begin{align}\label{eq:alpha_hat_def}
\hat \alpha_1=\Delta({\cal C}_{12}+{\cal E})-6{\cal B}_- -2{\cal G}{\cal A}^{(2)}, \qquad \hat \alpha_2={\cal E}-{\cal G}{\cal A }^{(1)}.
\end{align}
These are the strong field versions of the parameters $\alpha_1$ and $\alpha_2$, which are contained inside the definition of the calligraphic objects in \eqref{eq:alpha_hat_def}, so that $\hat \alpha_1$ and $\hat \alpha_2$ are implicit functions of them. They are directly proportional to each other only in the case in which the sensitivities vanish exactly.

In the absence of PN corrections, the motion of the two-body system describes a Keplerian orbit, parametrized by $\bm{x}=r\bm{n}$ with
\begin{align}
\nonumber \bm{n}&=[-\cos \Omega \cos(\omega+f)-\cos \iota \sin \Omega \sin(\omega+f)]\bm{e}_X+[\sin \Omega \cos(\omega+f) \\
&+\cos \iota \cos \Omega \sin (\omega+f) ]\bm{e}_Y+\sin \iota \sin (\omega+f) \bm{e}_Z,
\end{align}
where the orbital elements are: 
inclination $\iota$, longitude of the ascending node $\Omega$ and pericenter angle $\omega$. The element $f=\omega-\phi$ is the true anomaly, with $\phi$ the orbital phase measured from the ascending node. The reference vectors $\bm{e}_i$ form an orthonormal basis. 

When the extra force \eqref{eq:force} is included, Keplerian orbits are not solutions to the equations of motion anymore. However, provided that the force is small enough relative to the Newtonian force, we can use perturbation theory and translate the dependence on time of the motion to the orbital parameters. This is the method of \emph{osculating orbits} described in \cite{will_2018}, which leads to a secular variation of the orbital elements under the effect of $\bm{a}$. In order to parametrize this change in terms of the velocity vector of the preferred frame, we decompose the latter by projecting it onto the orbital plane by defining
\begin{align}
\omega_P=\bm{\omega}\cdot \bm{e}_P,\quad \omega_Q=\bm{\omega}\cdot\bm{e}_Q,\quad \omega_Z=\bm{\omega}\cdot\bm{z},
\end{align}
as well as onto the angular momentum vector
\begin{align}
    \omega_h=\bm{\omega}\cdot \bm{h}=\omega_Z\sqrt{{\cal G}mp},
\end{align}

Following the computation in \cite{will_2018}, we thus find that the local terms in $\bm{a}_{\rm L}$ induce a change only on the pericenter angle $\omega$, which in an orbit changes by
\begin{align}
\Delta_{\rm L} \omega=\frac{6\pi m}{{\cal G}p}\left[{\cal G}{\cal B}_++\frac{1}{6}\left({\cal G}^2-{\cal D}\right)+\frac{1}{6}{\cal G}\left(6 \Delta {\cal B}_- + \eta (2{\cal C}_{12}+{\cal E})+{\cal G}{\cal A}^{(3)}\right)\right],
\end{align}
and is of course independent of $\bm \omega$. 
Again, this agrees with \cite{will:2018ont} up to a typographical error in their result. The rest of secular changes vanish, either because they are identically zero or because they compensate along the orbit.

On the other hand, the force induced by preferred frame effects produces a secular change in all orbital parameters
\begin{align}
&\Delta_{\rm PF} a=\frac{2\pi e \omega_P}{(1-e^2)^2}\left( \frac{m p}{{\cal G}}\right)^{\frac{1}{2}}\left(\hat{\alpha}_1+4 {\cal A}^{(2)}{\cal G}\right),\label{eq:Delta_a}\\
&\Delta_{\rm PF} \iota =\pi \hat{\alpha}_1 \left(\frac{m}{{\cal G}p}\right)^{\frac{1}{2}} \omega_h \sin (\omega) e F(e) - \frac{2\pi \hat{\alpha}_2 \omega_h\omega_R F(e)}{{\cal G} \sqrt{1-e^2}},\\
&\Delta_{\rm PF} \Omega=-\pi \hat\alpha_1 \left(\frac{m}{{\cal G}p}\right)^{\frac{1}{2}} \frac{\omega_h}{\sin (\iota)}\cos(\omega) e F(e)-)\frac{2\pi \hat{\alpha}_2 \omega_h \omega_S F(e)}{{\cal G}\sin (\iota) \sqrt{1-e^2} },\\
\nonumber &\Delta_{\rm PF} \varpi =- \pi \hat \alpha_1 \left(\frac{m}{{\cal G} p}\right)^{\frac{1}{2}} \omega_Q \frac{\sqrt{1-e^2} F(e)}{e}-\pi \hat \alpha_2 (\omega_P^2 \omega_Q^2)F(e)^2\\ &\quad \qquad +  \frac{\pi\omega_Q}{e}\left(\frac{m}{{\cal G} p}\right)^{\frac{1}{2}}  (\hat{\alpha}_1+4 {\cal A}^{(2)}{\cal G}),\\
\nonumber &\Delta_{\rm PF} e= -\pi \hat \alpha_1 \left(\frac{m}{{\cal G}p}\right)^{\frac{1}{2}} \omega_P (1-e^2) F(e) + 2\pi \hat{\alpha}_2 \omega_P \omega_Q e \sqrt{1-e^2} F(e)^2\\
& \qquad \quad + \pi \omega_P \left(\frac{m}{{\cal G}p}\right)^{\frac{1}{2}} (\hat \alpha_1 + 4 {\cal A}^{(2)}{\cal G}),
\end{align}
where $\Delta_{\rm PF} \varpi = \Delta_{\rm PF}\omega+\cos(\iota)\Delta_{\rm PF}\Omega$ and
\begin{align}
&F(e)=\frac{1}{1+\sqrt{1-e^2}},\\
&\omega_R=\omega_P \cos\omega - \omega_Q \sqrt{1-e^2}\ \sin\omega,\\
&\omega_S=\omega_P \sin\omega + \omega_Q \sqrt{1-e^2}\ \cos\omega.
\end{align}

Out of these deviations, the most relevant one is the variation of the semimajor axis, which can be related to the change in the period of the orbit by using Kepler's third law
\begin{align}
\frac{\Delta T}{T}=\frac{3}{2}\frac{\Delta a}{a}.
\end{align}
Note however that this change is sub-leading with respect to the change expected from emission of gravitational radiation in a binary system like the one considered throughout this paper [c.f. Eq. \eqref{Pdot-AE}], which is actually the dominant factor.

\newpage
\section*{References}

\bibliographystyle{plain}
\bibliography{main}

\begin{thebibliography}{10}

\bibitem{Monitor:2017mdv}
B.~P. Abbott et~al.
\newblock {Gravitational Waves and Gamma-rays from a Binary Neutron Star
  Merger: GW170817 and GRB 170817A}.
\newblock {\em Astrophys. J.}, 848(2):L13, 2017.

\bibitem{PhysRevLett.119.161101}
B.~P. Abbott et~al.
\newblock Gw170817: Observation of gravitational waves from a binary neutron
  star inspiral.
\newblock {\em Phys. Rev. Lett.}, 119:161101, Oct 2017.

\bibitem{GBM:2017lvd}
B.~P. Abbott et~al.
\newblock {Multi-messenger Observations of a Binary Neutron Star Merger}.
\newblock {\em Astrophys. J.}, 848(2):L12, 2017.

\bibitem{ap4}
A.~Akmal, V.~R. Pandharipande, and D.~G. Ravenhall.
\newblock Equation of state of nucleon matter and neutron star structure.
\newblock {\em Phys. Rev. C}, 58:1804--1828, Sep 1998.

\bibitem{Anderson_2019}
David Anderson, Paulo Freire, and Nicolás Yunes.
\newblock Binary pulsar constraints on massless scalar–tensor theories using
  bayesian statistics.
\newblock {\em Classical and Quantum Gravity}, 36(22):225009, Oct 2019.

\bibitem{Antoniadis_2013}
J.~Antoniadis, P.~C.~C. Freire, N.~Wex, T.~M. Tauris, R.~S. Lynch, M.~H. van
  Kerkwijk, M.~Kramer, C.~Bassa, V.~S. Dhillon, T.~Driebe, and et~al.
\newblock A massive pulsar in a compact relativistic binary.
\newblock {\em Science}, 340(6131):1233232–1233232, Apr 2013.

\bibitem{archibald:2018oxs}
Anne~M. Archibald, Nina~V. Gusinskaia, Jason W.~T. Hessels, Adam~T. Deller,
  David~L. Kaplan, Duncan~R. Lorimer, Ryan~S. Lynch, Scott~M. Ransom, and
  Ingrid~H. Stairs.
\newblock {Universality of free fall from the orbital motion of a pulsar in a
  stellar triple system}.
\newblock {\em Nature}, 559(7712):73--76, 2018.

\bibitem{Audren:2014hza}
B.~Audren, D.~Blas, M.M. Ivanov, J.~Lesgourgues, and S.~Sibiryakov.
\newblock {Cosmological constraints on deviations from Lorentz invariance in
  gravity and dark matter}.
\newblock {\em JCAP}, 03:016, 2015.

\bibitem{Barausse:2019yuk}
Enrico Barausse.
\newblock {Neutron star sensitivities in Horava gravity after GW170817}.
\newblock {\em Phys. Rev. D}, 100(8):084053, 2019.

\bibitem{barvinsky:2015kil}
Andrei~O. Barvinsky, Diego Blas, Mario Herrero-Valea, Sergey~M. Sibiryakov, and
  Christian~F. Steinwachs.
\newblock {Renormalization of Horava gravity}.
\newblock {\em Phys. Rev.}, D93(6):064022, 2016.

\bibitem{barvinsky:2017kob}
Andrei~O. Barvinsky, Diego Blas, Mario Herrero-Valea, Sergey~M. Sibiryakov, and
  Christian~F. Steinwachs.
\newblock {Horava Gravity is Asymptotically Free in 2 + 1 Dimensions}.
\newblock {\em Phys. Rev. Lett.}, 119(21):211301, 2017.

\bibitem{Bednik:2013nxa}
Grigory Bednik, Oriol Pujolas, and Sergey Sibiryakov.
\newblock {Emergent Lorentz invariance from Strong Dynamics: Holographic
  examples}.
\newblock {\em JHEP}, 11:064, 2013.

\bibitem{Bettoni:2017lxf}
Dario Bettoni, Adi Nusser, Diego Blas, and Sergey Sibiryakov.
\newblock {Testing Lorentz invariance of dark matter with satellite galaxies}.
\newblock {\em JCAP}, 05:024, 2017.

\bibitem{1976ApJ...205..580B}
R.~{Blandford} and S.~A. {Teukolsky}.
\newblock {Arrival-time analysis for a pulsar in a binary system.}
\newblock {\em \apj}, 205:580--591, April 1976.

\bibitem{Blas:2009qj}
D.~Blas, O.~Pujolas, and S.~Sibiryakov.
\newblock {Consistent Extension of Horava Gravity}.
\newblock {\em Phys. Rev. Lett.}, 104:181302, 2010.

\bibitem{Blas:2012vn}
Diego Blas, Mikhail~M. Ivanov, and Sergey Sibiryakov.
\newblock {Testing Lorentz invariance of dark matter}.
\newblock {\em JCAP}, 10:057, 2012.

\bibitem{blas:2010hb}
Diego Blas, Oriol Pujolas, and Sergey Sibiryakov.
\newblock {Models of non-relativistic quantum gravity: The Good, the bad and
  the healthy}.
\newblock {\em JHEP}, 04:018, 2011.

\bibitem{Blas:2011zd}
Diego Blas and Hillary Sanctuary.
\newblock {Gravitational Radiation in Horava Gravity}.
\newblock {\em Phys. Rev.}, D84:064004, 2011.

\bibitem{Bonetti:2015oda}
Matteo Bonetti and Enrico Barausse.
\newblock {Post-Newtonian constraints on Lorentz-violating gravity theories
  with a MOND phenomenology}.
\newblock {\em Phys. Rev.}, D91:084053, 2015.
\newblock [Erratum: Phys. Rev.D93,029901(2016)].

\bibitem{Braun:2019gdo}
Robert Braun, Anna Bonaldi, Tyler Bourke, Evan Keane, and Jeff Wagg.
\newblock {Anticipated Performance of the Square Kilometre Array -- Phase 1
  (SKA1)}.
\newblock 12 2019.

\bibitem{10.1046/j.1365-8711.1998.01634.x}
Paul~J. Callanan, Peter~M. Garnavich, and Detlev Koester.
\newblock {The mass of the neutron star in the binary millisecond pulsar PSR
  J1012 + 5307}.
\newblock {\em Monthly Notices of the Royal Astronomical Society},
  298(1):207--211, 07 1998.

\bibitem{Carroll:2004ai}
Sean~M. Carroll and Eugene~A. Lim.
\newblock {Lorentz-violating vector fields slow the universe down}.
\newblock {\em Phys. Rev.}, D70:123525, 2004.

\bibitem{Chadha:1982qq}
S.~Chadha and Holger~Bech Nielsen.
\newblock {LORENTZ INVARIANCE AS A LOW-ENERGY PHENOMENON}.
\newblock {\em Nucl. Phys.}, B217:125--144, 1983.

\bibitem{Cornish:2017jml}
Neil Cornish, Diego Blas, and Germano Nardini.
\newblock {Bounding the speed of gravity with gravitational wave observations}.
\newblock {\em Phys. Rev. Lett.}, 119(16):161102, 2017.

\bibitem{Coughlin_2019}
Michael~W Coughlin, Tim Dietrich, Ben Margalit, and Brian~D Metzger.
\newblock Multimessenger bayesian parameter inference of a binary neutron star
  merger.
\newblock {\em Monthly Notices of the Royal Astronomical Society: Letters},
  489(1):L91–L96, Aug 2019.

\bibitem{0264-9381-9-9-015}
T~Damour and G~Esposito-Farese.
\newblock Tensor-multi-scalar theories of gravitation.
\newblock {\em Classical and Quantum Gravity}, 9(9):2093, 1992.

\bibitem{damour:1991rd}
Thibault Damour and Joseph~H. Taylor.
\newblock {Strong field tests of relativistic gravity and binary pulsars}.
\newblock {\em Phys. Rev.}, D45:1840--1868, 1992.

\bibitem{1975ApJ...196L..59E}
D.~M. {Eardley}.
\newblock {Observable effects of a scalar gravitational field in a binary
  pulsar}.
\newblock {\em {Astrophys. J. Lett.}}, 196:L59--L62, March 1975.

\bibitem{eling:2005zq}
Christopher Eling.
\newblock {Energy in the Einstein-aether theory}.
\newblock {\em Phys. Rev.}, D73:084026, 2006.
\newblock [Erratum: Phys. Rev.D80,129905(2009)].

\bibitem{Elliott:2005va}
Joshua~W. Elliott, Guy~D. Moore, and Horace Stoica.
\newblock {Constraining the new Aether: Gravitational Cerenkov radiation}.
\newblock {\em JHEP}, 08:066, 2005.

\bibitem{Gumrukcuoglu:2017ijh}
A.~Emir~Gumrukcuoglu, Mehdi Saravani, and Thomas~P. Sotiriou.
\newblock {Ho\v{r}ava gravity after GW170817}.
\newblock {\em Phys. Rev.}, D97(2):024032, 2018.

\bibitem{Foster:2006az}
Brendan~Z. Foster.
\newblock {Radiation damping in Einstein-aether theory}.
\newblock {\em Phys. Rev.}, D73:104012, 2006.
\newblock [Erratum: Phys. Rev.D75,129904(2007)].

\bibitem{foster:2007gr}
Brendan~Z. Foster.
\newblock {Strong field effects on binary systems in Einstein-aether theory}.
\newblock {\em Phys. Rev.}, D76:084033, 2007.

\bibitem{foster:2005dk}
Brendan~Z. Foster and Ted Jacobson.
\newblock {Post-Newtonian parameters and constraints on Einstein-aether
  theory}.
\newblock {\em Phys. Rev.}, D73:064015, 2006.

\bibitem{Franchini:2021bpt}
Nicola Franchini, Mario Herrero-Valea, and Enrico Barausse.
\newblock {On distinguishing between General Relativity and a class of
  khronometric theories}.
\newblock 3 2021.

\bibitem{Freire_2012}
Paulo C.~C. Freire, Norbert Wex, Gilles Esposito-Farèse, Joris P.~W. Verbiest,
  Matthew Bailes, Bryan~A. Jacoby, Michael Kramer, Ingrid~H. Stairs, John
  Antoniadis, and Gemma~H. Janssen.
\newblock The relativistic pulsar-white dwarf binary psr j1738+0333 - ii. the
  most stringent test of scalar-tensor gravity.
\newblock {\em Monthly Notices of the Royal Astronomical Society},
  423(4):3328–3343, Jun 2012.

\bibitem{Garfinkle:2011iw}
David Garfinkle and Ted Jacobson.
\newblock {A positive energy theorem for Einstein-aether and Ho\v{r}ava
  gravity}.
\newblock {\em Phys. Rev. Lett.}, 107:191102, 2011.

\bibitem{GrootNibbelink:2004za}
Stefan Groot~Nibbelink and Maxim Pospelov.
\newblock {Lorentz violation in supersymmetric field theories}.
\newblock {\em Phys. Rev. Lett.}, 94:081601, 2005.

\bibitem{horava:2009uw}
Petr Ho\v{r}ava.
\newblock {Quantum Gravity at a Lifshitz Point}.
\newblock {\em Phys. Rev.}, D79:084008, 2009.

\bibitem{hulse:1974eb}
R.~A. Hulse and J.~H. Taylor.
\newblock {Discovery of a pulsar in a binary system}.
\newblock {\em Astrophys. J.}, 195:L51--L53, 1975.

\bibitem{Jacobson:2004ts}
T.~Jacobson and D.~Mattingly.
\newblock {Einstein-Aether waves}.
\newblock {\em Phys. Rev.}, D70:024003, 2004.

\bibitem{Jacobson:2008aj}
Ted Jacobson.
\newblock {Einstein-aether gravity: A Status report}.
\newblock {\em PoS}, QG-PH:020, 2007.

\bibitem{Jacobson:2010mx}
Ted Jacobson.
\newblock {Extended Horava gravity and Einstein-aether theory}.
\newblock {\em Phys. Rev.}, D81:101502, 2010.
\newblock [Erratum: Phys. Rev.D82,129901(2010)].

\bibitem{jacobson_2014}
Ted Jacobson.
\newblock Undoing the twist: The hořava limit of einstein-aether theory.
\newblock {\em Physical Review D}, 89(8), Apr 2014.

\bibitem{Jacobson:2005bg}
Ted Jacobson, Stefano Liberati, and David Mattingly.
\newblock {Lorentz violation at high energy: Concepts, phenomena and
  astrophysical constraints}.
\newblock {\em Annals Phys.}, 321:150--196, 2006.

\bibitem{Jacobson:2000xp}
Ted Jacobson and David Mattingly.
\newblock {Gravity with a dynamical preferred frame}.
\newblock {\em Phys. Rev.}, D64:024028, 2001.

\bibitem{Jiang:2019vmf}
Nan Jiang and Kent Yagi.
\newblock {Improved Analytic Modeling of Neutron Star Interiors}.
\newblock {\em Phys. Rev.}, D99(12):124029, 2019.

\bibitem{kostelecky:2010ze}
Alan~V. Kostelecky and Jay~D. Tasson.
\newblock {Matter-gravity couplings and Lorentz violation}.
\newblock {\em Phys. Rev.}, D83:016013, 2011.

\bibitem{kostelecky:2003fs}
V.~Alan Kostelecky.
\newblock {Gravity, Lorentz violation, and the standard model}.
\newblock {\em Phys. Rev.}, D69:105009, 2004.

\bibitem{kostelecky:2008ts}
V.~Alan Kostelecky and Neil Russell.
\newblock {Data Tables for Lorentz and CPT Violation}.
\newblock {\em Rev. Mod. Phys.}, 83:11--31, 2011.

\bibitem{Kramer:2006nb}
M.~Kramer et~al.
\newblock {Tests of general relativity from timing the double pulsar}.
\newblock {\em Science}, 314:97--102, 2006.

\bibitem{10.1111/j.1365-2966.2009.15481.x}
K.~Lazaridis, N.~Wex, A.~Jessner, M.~Kramer, B.~W. Stappers, G.~H. Janssen,
  G.~Desvignes, M.~B. Purver, I.~Cognard, G.~Theureau, A.~G. Lyne, C.~A.
  Jordan, and J.~A. Zensus.
\newblock {Generic tests of the existence of the gravitational dipole radiation
  and the variation of the gravitational constant}.
\newblock {\em Monthly Notices of the Royal Astronomical Society},
  400(2):805--814, 11 2009.

\bibitem{liberati:2013xla}
Stefano Liberati.
\newblock {Tests of Lorentz invariance: a 2013 update}.
\newblock {\em Class.Quant.Grav.}, 30:133001, 2013.

\bibitem{Mattingly:2005re}
David Mattingly.
\newblock {Modern tests of Lorentz invariance}.
\newblock {\em Living Rev. Rel.}, 8:5, 2005.

\bibitem{muller:2005sr}
Jurgen Muller, James~G. Williams, and Slava~G. Turyshev.
\newblock {Lunar laser ranging contributions to relativity and geodesy}.
\newblock {\em Astrophys. Space Sci. Libr.}, 349:457--472, 2008.

\bibitem{Oost:2018tcv}
Jacob Oost, Shinji Mukohyama, and Anzhong Wang.
\newblock {Constraints on Einstein-aether theory after GW170817}.
\newblock {\em Phys. Rev.}, D97(12):124023, 2018.

\bibitem{Oppenheimer}
J.~R. Oppenheimer and G.~M. Volkoff.
\newblock On massive neutron cores.
\newblock {\em Phys. Rev.}, 55:374--381, Feb 1939.

\bibitem{Pospelov:2010mp}
Maxim Pospelov and Yanwen Shang.
\newblock {On Lorentz violation in Horava-Lifshitz type theories}.
\newblock {\em Phys. Rev.}, D85:105001, 2012.

\bibitem{ramos}
Oscar Ramos and Enrico Barausse.
\newblock {Constraints on Ho\v rava gravity from binary black hole
  observations}.
\newblock {\em Phys. Rev.}, D99(2):024034, 2019.

\bibitem{Sarbach:2019yso}
Olivier Sarbach, Enrico Barausse, and Jorge~A. Preciado-L\'opez.
\newblock {Well-posed Cauchy formulation for Einstein-\ae{}ther theory}.
\newblock {\em Class. Quant. Grav.}, 36(16):165007, 2019.

\bibitem{shao:2013wga}
Lijing Shao, R.~Nicolas Caballero, Michael Kramer, Norbert Wex, David~J.
  Champion, and Axel Jessner.
\newblock {A new limit on local Lorentz invariance violation of gravity from
  solitary pulsars}.
\newblock {\em Class. Quant. Grav.}, 30:165019, 2013.

\bibitem{shao:2012eg}
Lijing Shao and Norbert Wex.
\newblock {New tests of local Lorentz invariance of gravity with
  small-eccentricity binary pulsars}.
\newblock {\em Class. Quant. Grav.}, 29:215018, 2012.

\bibitem{taylor:1982zz}
J.~H. Taylor and J.~M. Weisberg.
\newblock {A new test of general relativity: Gravitational radiation and the
  binary pulsar PS R 1913+16}.
\newblock {\em Astrophys. J.}, 253:908--920, 1982.

\bibitem{Taylor:1989sw}
Joseph~H. Taylor and J.~M. Weisberg.
\newblock {Further experimental tests of relativistic gravity using the binary
  pulsar PSR 1913+16}.
\newblock {\em Astrophys. J.}, 345:434--450, 1989.

\bibitem{Thorne}
Kip~S. Thorne.
\newblock Multipole expansions of gravitational radiation.
\newblock {\em Rev. Mod. Phys.}, 52:299--339, Apr 1980.

\bibitem{tolman:1939jz}
Richard~C. Tolman.
\newblock {Static solutions of Einstein's field equations for spheres of
  fluid}.
\newblock {\em Phys. Rev.}, 55:364--373, 1939.

\bibitem{voisin_2020}
G.~Voisin, I.~Cognard, P.~C.~C. Freire, N.~Wex, L.~Guillemot, G.~Desvignes,
  M.~Kramer, and G.~Theureau.
\newblock An improved test of the strong equivalence principle with the pulsar
  in a triple star system.
\newblock {\em Astronomy \& Astrophysics}, 638:A24, Jun 2020.

\bibitem{Bull:2018lat}
A.~Weltman et~al.
\newblock {Fundamental physics with the Square Kilometre Array}.
\newblock {\em Publ. Astron. Soc. Austral.}, 37:e002, 2020.

\bibitem{Will:2014kxa}
Clifford~M. Will.
\newblock {The Confrontation between General Relativity and Experiment}.
\newblock {\em Living Rev. Rel.}, 17:4, 2014.

\bibitem{will:2018ont}
Clifford~M. Will.
\newblock {Testing general relativity with compact-body orbits: a modified
  Einstein\textendash{}Infeld\textendash{}Hoffmann framework}.
\newblock {\em Class. Quant. Grav.}, 35(8):085001, 2018.

\bibitem{will_2018}
Clifford~M. Will.
\newblock {\em Theory and Experiment in Gravitational Physics}.
\newblock Cambridge University Press, 2 edition, 2018.

\bibitem{Will:1989sk}
Clifford~M. Will and Helmut~W. Zaglauer.
\newblock {Gravitational Radiation, Close Binary Systems, and the Brans-dicke
  Theory of Gravity}.
\newblock {\em Astrophys. J.}, 346:366, 1989.

\bibitem{Yagi:2013ava}
Kent Yagi, Diego Blas, Enrico Barausse, and Nicolás Yunes.
\newblock {Constraints on Einstein-Æther theory and Hořava gravity from
  binary pulsar observations}.
\newblock {\em Phys. Rev.}, D89(8):084067, 2014.
\newblock [Erratum: Phys. Rev.D90,no.6,069902(2014); Erratum: Phys.
  Rev.D90,no.6,069901(2014)].

\bibitem{Yagi:2013qpa}
Kent Yagi, Diego Blas, Nicolas Yunes, and Enrico Barausse.
\newblock {Strong Binary Pulsar Constraints on Lorentz Violation in Gravity}.
\newblock {\em Phys. Rev. Lett.}, 112(16):161101, 2014.

\bibitem{Yagi:2013mbt}
Kent Yagi, Leo~C. Stein, Nicolas Yunes, and Takahiro Tanaka.
\newblock {Isolated and Binary Neutron Stars in Dynamical Chern-Simons
  Gravity}.
\newblock {\em Phys. Rev. D}, 87:084058, 2013.
\newblock [Erratum: Phys.Rev.D 93, 089909 (2016)].

\bibitem{PhysRevD.100.083012}
Xiang Zhao, Chao Zhang, Kai Lin, Tan Liu, Rui Niu, Bin Wang, Shaojun Zhang,
  Xing Zhang, Wen Zhao, Tao Zhu, and Anzhong Wang.
\newblock Gravitational waveforms and radiation powers of the triple system psr
  $\mathrm{J}0337+1715$ in modified theories of gravity.
\newblock {\em Phys. Rev. D}, 100:083012, Oct 2019.

\end{thebibliography}
\end{document}